\documentclass[acmlarge]{acmart}

\usepackage{booktabs} 
\usepackage{graphicx}
\usepackage{url}
\usepackage{amsmath}
\usepackage[compatibility=false]{caption}
\usepackage{subcaption}
\usepackage{array}
\usepackage{lineno}

\usepackage[ruled]{algorithm2e} 

\SetAlFnt{\small}
\SetAlCapFnt{\small}
\SetAlCapNameFnt{\small}
\SetAlCapHSkip{0pt}
\IncMargin{-\parindent}

\setcopyright{rightsretained}

\acmDOI{0000001.0000001}

\begin{document}

\title{Entity-Relationship Search over the Web}

\acmConference[]{}{October 2018}{}
\acmYear{2018}
\copyrightyear{2018}

\author{Pedro Saleiro}
\affiliation{%
  \institution{Department of Computer Science, University of Chicago}
}
  \email{saleiro@uchicago.edu}

\author{Natasa Milic-Frayling}
\affiliation{%
  \institution{School of Computer Science, University of Nottingham}
}
\email{natasa.milic-frayling@nottingham.ac.uk }

\author{Eduarda Mendes Rodrigues}
\affiliation{%
  \institution{Faculdade de Engenharia, Universidade do Porto}
}
\email{eduarda@fe.up.pt}

\author{Carlos Soares}
\affiliation{%
  \institution{Faculdade de Engenharia, Universidade do Porto}
}
\email{csoares@fe.up.pt}


\begin{abstract}

Entity-Relationship (E-R) Search is a complex case of Entity Search where the goal is to search for multiple unknown entities and relationships connecting them. We assume that a E-R query can be decomposed as a sequence of sub-queries each containing keywords related to a specific entity or relationship. We adopt a probabilistic formulation of the E-R search problem. When creating specific representations for entities (e.g. context terms) and for pairs of entities (i.e. relationships) it is possible to create a graph of probabilistic dependencies between sub-queries and entity plus relationship representations. To the best of our knowledge this represents the first probabilistic model of E-R search. We propose and develop a novel supervised Early Fusion-based model for E-R search, the Entity-Relationship Dependence Model (ERDM). It uses Markov Random Field to model term dependencies of E-R sub-queries and entity/relationship documents.  We performed experiments with more than 800M entities and relationships extractions from ClueWeb-09-B with FACC1 entity linking. We obtained promising results using 3 different query collections comprising 469 E-R queries, with results showing that it is possible to perform E-R search without using fix and pre-defined entity and relationship types, enabling a wide range of queries to be addressed.
\end{abstract}

\maketitle

\section{Introduction}

In recent years, we have seen increased interest in using online information sources to find concise and precise information about specific issues, events, and entities rather than retrieving and reading entire documents and web pages. Modern search engines are now presenting entity cards, summarizing entity properties and related entities, to answer entity-bearing queries directly in the search engine result page. Examples of such queries are ``Who founded Intel?" and ``Works by Charles Rennie Mackintosh". 

Existing strategies for entity search can be divided in IR-centric and Semantic-Web-based approaches. The former usually rely on statistical language models to match and rank co-occurring terms in the proximity of the target entity \cite{balog2012expertise}. The latter consists in creating a SPARQL query and using it over a structured knowledge base to retrieve relevant RDF triples \cite{heath2011linked}. Neither of these paradigms provide good support for entity-relationship (E-R) retrieval, i.e., searching for multiple unknown entities and relationships connecting them. 

Contrary to traditional entity queries, E-R queries expect tuples of connected entities as answers. For instance, ``US technology companies contracts Chinese electronics manufacturers" can be answered by tuples $<$\textit{Apple}, \textit{Foxconn}$>$, while ``Companies founded by disgraced Hollywood producer" is expecting tuples $<$\textit{Miramax}, \textit{Harvey Weinstein}$>$. In essence, an E-R query can be decomposed into a set of sub-queries that specify types of entities and types of relationships between entities.

Recent work in Semantic-Web search tackled E-R retrieval by extending SPARQL to support joins of multiple query results and creating an extended knowledge graph \cite{yahya2016relationship}. Extracted entities and relationships are typically stored in a knowledge graph. However, it is not always convenient to rely on a structured knowledge graph with predefined and constraining entity types. 

In particular, we are interested in transient information sources, such as online news or social media. General purpose knowledge graphs are usually fed with more stable and reliable data sources (e.g. Wikipedia). Furthermore, predefining and constraining entity and relationship types, such as in Semantic Web-based approaches, reduces the range of queries that can be answered and therefore limits the usefulness of entity search, particularly when one wants to leverage free-text.

To the best of our knowledge, E-R retrieval using IR-centric approaches is a new and unexplored research problem within the Information Retrieval research community. One of the objectives of our research is to explore to what degree we can leverage the textual context of entities and relationships, i.e., co-occurring terminology, to relax the notion of an entity or relationship type. 

Instead of being characterized by a fixed type, e.g., \textit{person}, \textit{country}, \textit{place}, the entity would be characterized by any contextual term. The same applies to the relationships. Traditional knowledge graphs have fixed schema of relationships, e.g. \textit{child of}, \textit{created by}, \textit{works for} while our approach relies on contextual terms in the text proximity of every two co-occurring entities in a raw document. Relationships descriptions such as ``criticizes'', ``hits back'', ``meets'' or ``interested in'' would be possible to search for. This is expected to significantly reduce the limitations which structured approaches suffer from, enabling a wider range of queries to be addressed. 

We assume that a E-R query can be formulated as a sequence of individual sub-queries each targeting a specific entity or relationship. If we create specific representations for entities (e.g. context terms) as well as for pairs of entities, i.e. relationships then we can create a graph of probabilistic dependencies between sub-queries and entity/relationship representations. We show that this dependencies can be depicted in a probabilistic graphical model, i.e. a Bayesian network. Therefore, answering an E-R query can be reduced to a computation of factorized conditional probabilities over a graph of sub-queries and entity/relationship documents. 

However, it is not possible to compute these conditional probabilities directly from raw documents in a collection. Such as with traditional entity retrieval, documents serve as proxies to entities (and relationships) representations. It is necessary to fuse information spread across multiple documents. We propose an early fusion approach to create entity and relationship centric document representations. It consists in aggregating context terms of entity and relationship occurrences to create two dedicate indexes, the \textit{entity} index and the \textit{relationship} index. Then it is possible to use any retrieval method to compute the relevance score of entity and relationship documents given the E-R sub-queries. Furthermore, we propose the Entity-Relationship Dependence Model, a novel early-fusion supervised model based on the Markov Random Field framework for Retrieval. We performed experiments at scale with results showing that it is possible to perform E-R retrieval without using fix and pre-defined entity and relationship types, enabling a wide range of queries to be addressed

\section{Related Work}

\subsection{Entity and Relationship Search}
Entity Search differs from traditional document search in the search unit. While document search considers a document as the atomic response to a query, in Entity Search document boundaries are not so important and entities need to be identified based on occurrence in documents \cite{adafre2007entity}. The focus level is more granular as the objective is to search and rank entities among documents. However, traditional Entity Search do not exploit semantic relationships between terms in the query and in the collection of documents, i.e. if there is no match between query terms and terms describing the entity, relevant entities tend to be missed.
 
Entity Search has been an active research topic in the last decade, including various specialized tracks, such as Expert finding track \cite{chen2006social}, INEX entity ranking track \cite{demartini2009overview}, TREC entity track \cite{balog2010overview} and SIGIR EOS workshop \cite{balog2012first}. Previous research faced two major challenges: entity representation and entity ranking. Entities are complex objects composed by a different number of properties and they are mentioned in a variety of contexts through time. Consequently, there is no single definition of the atomic unit (entity) to be retrieved. Additionally, it is a challenge to devise entity rankings that use various entity representations approaches and tackle different information needs. 

There are two main approaches for tackling Entity Search: \textit{Model 1}  or ``profile based approach'' and  \textit{Model 2} or ``voting approach'' \cite{balog2006formal}). The ``profile based approach'' starts by applying NER and NED in the collection in order to extract all entities occurrences. Then, for each entity identified, a meta-document is created by concatenating every passage in which the entity occurs. An index of entity meta-documents is created and a standard document ranking method (e.g. BM25) is applied to rank meta-documents with respect to a given query \cite{azzopardi2005language, craswell2005overview}. One of the main challenges of this approach is the transformation of original text documents to an entity-centric meta-document index, including pre-processing the collection in order to extract all entities and their context.

In the `voting approach'', the query is processed as typical document search to obtain an initial list of documents  \cite{balog2006formal, ru2005trec}. Entities are extracted from these documents using NER and NED techniques. Then, score functions are calculated to estimate the relation of entities captured and the initial query. For instance, counting the frequency of occurrence of the entity in the top documents combined with each document score (relevance to the query)  \cite{balog2006formal}. Another approach consists in taking into account the distance between the entity mention and the query terms in the documents \cite{petkova2007proximity}.

Recently, there is an increasing research interest in Entity Search over Linked Data, also referred as Semantic Search, due to the availability of structured information about entities and relations in the form of Knowledge Bases \cite{bron2013example,zong2015discovering,zhiltsov2015fielded}. Semantic Search exploit rich structured entity related in machine readable RDF format, expressed as a triple (entity, predicate, object). There are two types of search: keyword-based and natural language based search \cite{pound2012interpreting,unger2012template}. Regardless of the search type, the objective is to interpret the semantic structure of queries and translate it to the underlying schema of the target Knowledge Base. Most of the research focus is on interpreting the query intent \cite{pound2012interpreting,unger2012template} while others focus on how to  devise a ranking framework that deals with similarities between different attributes of the entity entry in the KB and the query terms \cite{zhiltsov2015fielded}

 Li et al. \cite{li2012entity} were the first to study relationship queries for structured querying entities over Wikipedia text with multiple predicates. This work used a query language with typed variables, for both entities and entity pairs, that integrates text conditions. First it computes individual predicates and then aggregates multiple predicate scores into a result score. The proposed method to score predicates relies on redundant co-occurrence contexts.

Yahya et al. \cite{yahya2016relationship} defined relationship queries as SPARQL-like subject-predicate-object (SPO) queries joined by one or more relationships. Authors cast this problem into a structured query language (SPARQL) and extended -it to support textual phrases for each of the SPO arguments. Therefore it allows to combine both structured SPARQL-like triples and text simultaneously. TriniTs extended the YAGO knowledge base with triples extracted from ClueWeb using an Open Information Extraction approach \cite{schmitz2012open}.

In the scope of relational databases, keyword-based graph search has been widely studied, including ranking \cite{yu2009keyword}. However, these approaches do not consider full documents of graph nodes and are limited to structured data. While searching over structured data is precise it can be limited in various respects. In order to increase the recall when no results are returned and enable prioritization of results when there are too many, Elbassuoni et al. \cite{elbassuoni2009language} propose a language-model for ranking results. Similarly, the models like EntityRank by Cheng et al. \cite{cheng2007entityrank} and Shallow Semantic Queries by Li et al. \cite{li2012entity}, relax the predicate definitions in the structured queries and, instead, implement proximity operators to bind the instances across entity types. Yahya et al. \cite{yahya2016relationship} propose algorithms for application of a set of relaxation rules that yield higher recall.

Web documents contain term information that can be used to apply pattern heuristics and statistical analysis often used to infer entities as investigated by Conrad and Utt \cite{conrad1994system}, Petkova and Croft \cite{petkova2007proximity}, Rennie and Jaakkola \cite{rennie2005using}. In fact, early work by Conrad and Utt \cite{conrad1994system} demonstrates a method that retrieves entities located in the proximity of a given keyword. They show that by using a fixed-size window around proper-names can be effective for supporting search for people and finding relationship among entities. Similar considerations of the co-occurrence statistics have been used to identify salient terminology, i.e. keyword to include in the document index \cite{petkova2007proximity}.

Existing approaches to the problem of entity-relationship (E-R) search are limited by pre-defined sets of both entity and relationship types. In this work, we generalize the problem to allow the search for entities and relationships without any restriction to a given set and we propose an IR-centric approach to address it.

\subsection{Markov Random Field for Retrieval}
The Markov Random Field (MRF) model for retrieval was first proposed by Metzler and Croft \cite{metzler2005markov} to model query term and document dependencies. 
In the context of retrieval, the objective is to rank documents by computing the posterior $P(D|Q)$, given a document $D$ and a query $Q$:

\begin{equation}\label{eq:one}
P(D|Q) =  \frac{P(Q,D)}{P(Q)} 
\end{equation}

For that purpose, a MRF is constructed from a graph $G$, which follows the local Markov property: every random variable in $G$ is independent of its non-neighbors given observed values for its neighbors. Therefore, different edge configurations imply different independence assumptions. 
\begin{figure}[h] 
\centering
\includegraphics[width=0.8\linewidth]{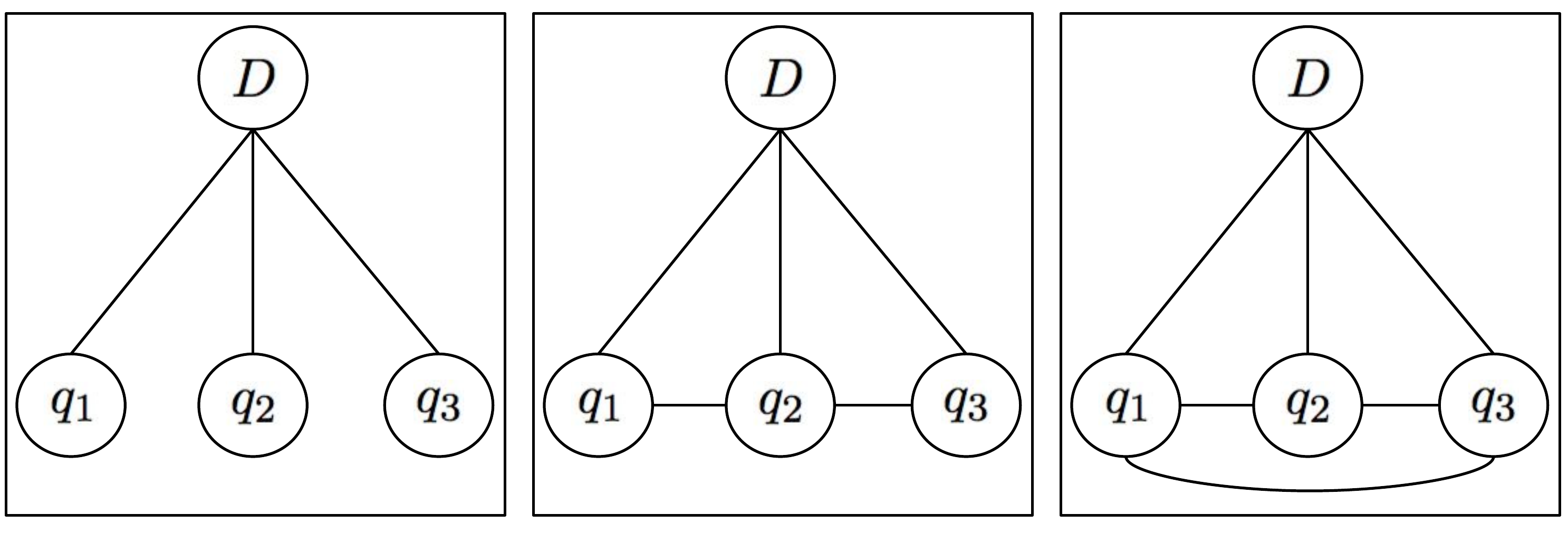}
\caption{Markov Random Field document and term dependencies.}\label{mrf}
\end{figure}

Metzler and Croft \cite{metzler2005markov} defined that $G$ consists of query term nodes $q_i$ and a document node $D$, as depicted in Figure \ref{mrf}. The joint probability mass function over the random variables in $G$ is defined by:

\begin{equation} \label{eq:two}
P_{G, \Lambda} (Q,D) = \frac{1}{Z_ \Lambda } \prod_{c \in C(G)} \psi(c;\Lambda)
\end{equation}

where $Q=q_1,...q_n$ are the query term nodes, $D$ is the document node, $C(G)$ is the set of  maximal cliques in $G$, and $\psi(c;\Lambda)$ is a non-negative potential function over clique configurations. The parameter $Z_ \Lambda = \sum_{Q,D}\prod_{c \in C(G)} \psi(c;\Lambda)$ is the partition function that normalizes the distribution. It is generally unfeasible to compute $Z_ \Lambda$, due to the exponential number of terms in the summation, and it is ignored as it does not influence ranking. 

The potential functions are defined as compatibility functions between nodes in a clique. For instance, a tf-idf score can be measured to reflect the ``aboutness'' between a query term $q_i$ and a document $D$. Metzler and Croft \cite{metzler2005markov} propose to associate one or more real valued feature function with each clique in the graph. The non-negative potential functions are defined using an exponential form $\psi(c;\Lambda) = \text{exp} [\lambda_c f(c)]$, where $\lambda_c$  is a feature weight, which is a free parameter in the model, associated with feature function $f(c)$. The model allows parameter and feature functions sharing across cliques of the same configuration, i.e. same size and type of nodes (e.g. 2-cliques of one query term node and one document node).

For each query $Q$, we construct a graph representing the query term dependencies, define a set of non-negative potential functions over the cliques of this graph and rank documents in descending order of $P_{\Lambda}(D|Q)$:

\begin{align}
\begin{split}\label{eq:3}
   P_{\Lambda}(D|Q)  {} & \stackrel{\text{$rank$}}{=}  \text{log } P_{\Lambda}(D|Q)\\
    & \stackrel{\text{$rank$}}{=} \text{log } P_{\Lambda}(Q,D) - \text{log } P_{\Lambda}(Q)\\
	& \stackrel{\text{$rank$}}{=}  \sum_{c \in C(G)} \text{log } \psi(c;\Lambda)\\
    & \stackrel{\text{$rank$}}{=}  \sum_{c \in C(G)} \text{log } \text{exp} [\lambda_c f(c)]\\
	& \stackrel{\text{$rank$}}{=}  \sum_{c \in C(G)} \lambda_c f(c)\\
\end{split}\\
\end{align}

Metzler and Croft concluded that given its general form, the MRF can emulate most of the retrieval and dependence models, such as language models \cite{song1999general}.

\subsubsection{Sequential Dependence Model}

The Sequential Dependence Model (SDM) is the most popular variant of the MRF retrieval model  \cite{metzler2005markov}. It defines two clique configurations represented in the following potential functions $ \psi (q_i, D;  \Lambda )$ and $\psi (q_i, q_{i+1}, D;  \Lambda )$. Basically, it considers sequential dependency between adjacent query terms and the document node.

The potential function of the 2-cliques containing a query term node and a document node is represented as $ \psi (q_i, D;  \Lambda ) =  \text{exp} [ \lambda_T f_T(q_i, D)]$. The clique configuration containing contiguous query terms and a document node is represented by two real valued functions. The first considers exact ordered matches of the two query terms in the document, while the second aims to capture unordered matches within $N$ fixed window sizes. Consequently, the second potential function is $\psi (q_i, q_{i+1}, D;  \Lambda ) = \text{exp} [ \lambda_O f_O(q_i,q_{i+1},D) + \lambda_U f_U(q_i,q_{i+1},D)]$.

Replacing $\psi(c;\Lambda)$ by these potential functions in Equation \ref{eq:3} and factoring out the parameters $\lambda$, the SDM can be represented as a mixture model computed over term, phrase and proximity feature classes:

\begin{flalign*}
P(D|Q)   \stackrel{\text{$rank$}}{=} {} &  \lambda_T \text{ } \text{ } \text{ } \text{ } \sum_{q_i \in Q} \text{ } \text{ }  \text{ } f_T(q_i, D) \text{ } + \\
& \lambda_O \sum_{q_i,q_{i+1}  \in Q} f_O(q_i,q_{i+1},D) \text{ } + \\
& \lambda_U  \sum_{q_i,q_{i+1}  \in Q} f_U(q_i,q_{i+1},D)\\
\end{flalign*}

where the free parameters $\lambda$ must follow the constraint $\lambda_T + \lambda_O + \lambda_U = 1$. Coordinate Ascent was chosen to learn the optimal $\lambda$ values that maximize mean average precision using training data \cite{metzler2007linear}.
Considering $tf$ the frequency of the term(s) in the document $D$, $cf$ the frequency of the term(s) in the entire collection $C$, the feature functions in SDM are set as:

\begin{equation} \label{eq:sdm_t}
 f_T(q_i,D) =  \text{log} \left [\frac{ tf_{q_i,D} + \mu \frac{ cf_{q_i}}{|C|}}{|D| + \mu} \right ]
\end{equation}

\begin{equation} \label{eq:sdm_o}
f_O(q_i,q_{i+1},D) =  \text{log} \left [\frac{tf_{\#1(q_i,q_{i+1}),D}  + \mu \frac{ cf_{\#1(q_i,q_{i+1)}}}{|C|}}{|D| + \mu} \right ]
\end{equation}

\begin{equation} \label{eq:sdm_u}
f_U(q_i,q_{i+1},D)=    \text{log} \left [\frac{tf_{\#uwN(q_i,q_{i+1}),D}  + \mu \frac{ cf_{\#uwN(q_i,q_{i+1)}}}{|C|}}{|D| + \mu} \right ]
\end{equation}

where $\mu$ is the Dirichlet prior for smoothing, $\#1(q_i,q_{i+1})$ is a function that searches for exact matches of the phrase ``$q_i$ $ q_{i+1}$'' and $\#uwN(q_i,q_{i+1})$ is a function that searches for co-occurrences of $q_i$ and $q_{i+1}$ within a window of fixed-N terms (usually 8 terms) across document $D$. SDM has shown state-of-the-art performance in ad-hoc document retrieval when compared with several bigram dependence models and standard bag-of-words retrieval models, across short and long queries \cite{huston2014comparison}.

\subsubsection{MRF for Entity Retrieval}
The current state-of-the-art methods in ad-hoc entity retrieval from knowledge graphs are based on MRF \cite{zhiltsov2015fielded,nikolaev2016parameterized}. The Fielded Sequential Dependence Model (FSDM) \cite{zhiltsov2015fielded} extends SDM for structured document retrieval and it is applied to entity retrieval from knowledge graphs. In this context, entity documents are composed by fields representing metadata about the entity. Each entity document has five fields: names, attributes, categories, similar entity names and related entity names. FSDM builds individual language models for each field in the knowledge base. This corresponds to replacing SDM feature functions with those of the Mixture of Language Models  \cite{ogilvie2003combining}. The feature functions of FSDM are defined as:

\begin{equation} \label{eq:fsdm_t}
 \tilde{f}_T(q_i,D) =  \text{log} \sum_{j}^{F}  w_{j}^{T} \left [\frac{ tf_{q_i,D_j} + \mu_j \frac{ cf_{q_i,j}}{|C_j|}}{|D_j| + \mu_j} \right ]
\end{equation}

\begin{equation} \label{eq:fsdm_o}
\tilde{f}_O(q_i,q_{i+1},D) =  \text{log} \sum_{j}^{F}  w_{j}^{O}   \left [\frac{tf_{\#1(q_i,q_{i+1}),D_j}  + \mu_j \frac{ cf_{\#1(q_i,q_{i+1}),j}}{|C_j|}}{|D_j| + \mu_j} \right ]
\end{equation}

\begin{equation} \label{eq:fsdm_u}
 \tilde{f}_U(q_i,q_{i+1},D)=    \text{log}  \sum_{j}^{F}  w_{j}^{U}    \left [\frac{tf_{\#uwN(q_i,q_{i+1}),D_j}  + \mu_j \frac{ cf_{\#uwN(q_i,q_{i+1}),j}  }{|C_j|}}{|D_j| + \mu_j} \right ]
\end{equation}

where $\mu_j$ are the Dirichlet priors for each field and $w_j$ are the weights for each field and must be non-negative with constraint $\sum_{j}^{F} w_j= 1$. Coordinate Ascent was used in two stages to learn $w_j$ and $\lambda$ values \cite{zhiltsov2015fielded}.

The Parameterized Fielded Sequential Dependence Model (PFSDM) \cite{nikolaev2016parameterized} extends the FSDM by dynamically calculating the field weights $w_j$ to different query terms. Part-of-speech features are applied to capture the relevance of query terms to specific fields of entity documents. For instance, \textit{NNP} feature is positive if query terms are proper nouns, therefore the query terms should be mapped to the \textit{names} field. Therefore, the field weight contribution of a given query term $q_i$ and a query bigram $q_i$,$q_{i+1}$ in a field $j$ are a linear weighted combination of features:

\begin{equation}
w_{q_{i},j} = \sum_{k} \alpha_{j,k}^{U} \phi_k (q_i, j)
\end{equation}
\begin{equation}
w_{q_{i},q_{i+1},j} = \sum_k \alpha_{j,k}^{B} \phi_k (q_{i},q_{i+1}, j)
\end{equation}

where $\phi_k(q_i, j)$ is the $k$ feature function of a query unigram for the field $j$ and $\alpha_{j,k}^{U}$ is its respective weight. For bigrams, 
$\phi_k (q_{i},q_{i+1}, j)$ is the $k$ feature function of a query bigram for the field $j$ and $\alpha_{j,k}^{B}$ is its respective weight.
Consequently, PFSDM has $F * U + F * B + 3$ total parameters, where $F$ is the number of fields, $U$ is the number of field mapping features for unigrams, $B$ is the number of field mapping features for bigrams, plus the three $\lambda$ parameters. Their estimation is performed in a two stage optimization. First $\alpha$ parameters are learned separately for unigrams and then bigrams. This is achieved  by setting to zero the corresponding $\lambda$ parameters. In the second stage, the $\lambda$ parameters are learned. Coordinate Ascent is used in both stages.

 The ELR model exploits entity mentions in queries by defining a dependency between entity documents and entity links in the query \cite{hasibi2016exploiting}.

\section{Modeling Entity-Relationship Retrieval}
 
E-R retrieval is a complex case of entity retrieval. E-R queries expect tuples of related entities as results instead of a single ranked list of entities as it happens with general entity queries. For instance, the E-R query ``Ethnic groups by country" is expecting a ranked list of tuples $<$\textit{ethnic group}, \textit{country}$>$ as results.  The goal is to search for multiple unknown entities and relationships connecting them.

\begin{table}[h]
\centering
\caption{E-R retrieval definitions.}
\label{tab:er_definitions}
\begin{tabular}{|p{1cm}|p{8cm}|}
\hline
$Q$ &  E-R query (e.g. ``congresswoman hits back at US president'').  \\ \hline
$Q^{E_i}$ &    Entity sub-query in $Q$  (e.g. ``congresswoman'').  \\ \hline
$Q^{R_{i-1,i}}$ &   Relationship sub-query in $Q$ (e.g. ``hits back at'').  \\ \hline
$D^{E_i}$& Term-based representation of an entity (e.g. <Frederica Wilson> = $\{$representative, congresswoman$\}$). We use the terminology \textit{representation} and \textit{document} interchangeably.  \\ \hline
$D^{R_{i-1,i}}$ &  Term-based representation of a relationship (e.g. <Frederica Wilson, Donald Trump> = $\{$hits,back$\}$). We use the terminology \textit{representation} and  \textit{document} interchangeably.  \\ \hline
$Q^{E}$ & The set of entity sub-queries in a E-R query (e.g. $\{$``congresswoman'',``US president'' $\}$). \\ \hline
$Q^{R}$ & The set of relationship sub-queries in a E-R query. \\ \hline
$D^{E}$ &  The set of entity documents to be retrieved by a E-R query. \\ \hline
$D^{R}$ &  The set of relationship documents to be retrieved by a E-R query. \\ \hline
$|Q|$ &   E-R query length corresponding to the number of entity and relationship sub-queries. \\ \hline
$T_E$ &   The entity tuple to be retrieved (e.g. <Frederica Wilson, Donald Trump>).  \\ \hline
\end{tabular}
\end{table}

In this section, we present a definition of E-R queries and a probabilistic formulation of the E-R retrieval problem from an Information Retrieval perspective. Table \ref{tab:er_definitions} presents several definitions that will be used throughout this chapter.

\subsection{E-R Queries} \label{erqueries}

E-R queries aim to obtain a ordered list of entity tuples $T_E=$ $<$$E_1, E_{2},..., E_n$$>$ as a result. Contrary to entity search queries where the expected result is a ranked list of single entities, results of E-R queries should contain two or more entities. For instance, the complex information need ``\textit{Silicon Valley companies founded by Harvard graduates}'' expects entity-pairs (2-tuples) $<$\textit{company}, \textit{founder}$>$ as results. In turn, ``\textit{European football clubs in which a Brazilian player won a trophy}" expects triples (3-tuples) $<$\textit{club}, \textit{player}, \textit{trophy}$>$ as results. 

Each pair of entities $E_{i-1}$, $E_{i}$ in an entity tuple is connected with a relationship $R(E_{i-1},E_i)$. A complex information need can be expressed in a relational format, which is decomposed into a set of sub-queries that specify types of entities $E$ and types of relationships $R(E_{i-1},E_{i})$ between entities. 

For each relationship sub-query there must be two sub-queries, one for each of the entities involved in the relationship. Thus a E-R query $Q$ that expects 2-tuples, is mapped into a triple of sub-queries $Q=\{Q^{E_1}$, $Q^{R_{1,2}}$, $Q^{E_{2}}\}$, where $Q^{E_1}$ and $Q^{E_{2}}$ are the entity attributes queried for $E_1$ and $E_{2}$ respectively, and $Q^{R_{1,2}}$ is a relationship attribute describing $R(E_i,E_{i+1})$. 

If we consider a E-R query as a chain of entity and relationship sub-queries $Q = \{Q^{E_1}$, $Q^{R_{1,2}}$, $Q^{E_{2}}$, ..., $Q^{E_{n-1}}$,$Q^{R_{n-1,n}}$, $Q^{E_n}  \}$ and we define the length of a E-R query $|Q|$ as the number of sub-queries, then the number of entity sub-queries must be $\frac{|Q|+1}{2}$ and the number of relationship sub-queries equal to $\frac{|Q|-1}{2}$. Consequently, the size of each entity tuple $T_E$ to be retrieved must be equal to the number of entity sub-queries. For instance, the E-R query ``soccer players who dated a top model'' with answers such as $<$\textit{Cristiano Ronaldo}, \textit{Irina Shayk}$>$) is represented as three sub-queries $Q^{E_1}=\{$\textit{soccer players}$\}$, $Q^{R_{1,2}}=\{$\textit{dated}$\}$, $Q^{E_{2}}=\{$\textit{top model}$\}$. 

Automatic mapping of terms from a E-R query $Q$ to sub-queries $Q^{E_i}$ or $Q^{R_{i-1,i}}$ is out of the scope of this work and can be seen as a problem of query understanding \cite{yahya2012natural,pound2012interpreting,sawant2013learning}. We assume that the information needs are decomposed into constituent entity and relationship sub-queries using Natural Language Processing techniques or by user input through an interface that enforces the structure $Q = \{Q^{E_1}$, $Q^{R_{1,2}}$, $Q^{E_{2}}$, ..., $Q^{E_{n-1}}$,$Q^{R_{n-1,n}}$, $Q^{E_n}  \}$.

\subsection{Bayesian E-R Retrieval}

Our approach to E-R retrieval assumes that we have a raw document collection (e.g. news articles) and each document $D_j$ is associated with one or more entities $E_i$. In other words, documents contain mentions to one or more entities that can be related between them. Since our goal is to retrieve tuples of related entities given a E-R query that expresses entity attributes and relationship attributes, we need to create term-based representations for both entities and relationships. We denote a representation of an entity $E_i$ as $D^{E_i}$. 

In E-R retrieval we are interested in retrieving tuples of entities $T_E=$ $<$$E_1, E_{2},..., E_n$$>$ as a result. The number of entities in each tuple can be two, three or more depending on the structure of the particular E-R query. When a E-R query aims to get tuples of more than two entities, we assume it is possible to combine tuples of length two. For instance, we can associate two tuples of length two that share the same entity to retrieve a tuple of length three. Therefore we create representations of relationships as pairs of entities. We denote a representation of a relationship $R(E_{i-1}, E_i)$ as $D^{R_{i-1,i}}$. 

Considering the example query ``Which spiritual leader won the same award as a US vice president?'' it can be formulated in the relational format as $Q^{E_1}= \{$\textit{spiritual leader}$\}$, $Q^{R_{1,2}}=\{$\textit{won}$\}$, $Q^{E_{2}}=\{$\textit{award}$\}$, $Q^{R_{2,3}}=\{$\textit{won}$\}$, $Q^{E_{3}}=\{$\textit{US vice president}$\}$. Associating the tuples of length two $<$Dalai Lama, Nobel Peace Prize$>$ and $<$Nobel Peace Prize, Al Gore$>$ would result in the expected 3-tuple $<$Dalai Lama, Nobel Peace Prize, Al Gore$>$. 

For the sake of clarity we now consider an example E-R query with three sub-queries ($|Q|=3$). This query aims  to retrieve a tuple of length two, i.e. a pair of entities connected by a relationship. Based on the definition of a E-R query, each entity in the resulting tuple must be relevant to the corresponding entity sub-queries $Q^E$. Moreover, the relationship between the two entities must also be relevant to the relationship sub-queries $Q^R$. Instead of calculating a simple posterior $P(D|Q)$ as with traditional information retrieval, in E-R retrieval the objective is to rank tuples based on a joint posterior of multiple entity and relationship representations given a E-R query, such as $P(D^{E_{2}},D^{E_1},D^{R_{1,2}}|Q)$ when $|Q|=3$.  

E-R queries can be seen as chains of interleaved entity and relationship sub-queries. We take advantage of the chain rule to formulate the joint probability  $P(D^{E_2}, D^{E_{1}},D^{R_{1,2}},Q)$ as a product of conditional probabilities. Formally, we want to rank entity and relationship candidates in descending order of the joint posterior $P(D^{E_2}, D^{E_{1}},D^{R_{1,2}}|Q)$ as:

\begin{align}
\begin{split}\label{erprob}
  P(D^{E_2}, D^{E_{1}},D^{R_{1,2}}|Q)  & \stackrel{\text{$rank$}}{=}   \frac{P(D^{E_2}, D^{E_{1}},D^{R_{1,2}},Q)}{P(Q)}\\
	& \stackrel{\text{$rank$}}{=}   \frac{P(D^{E_2}| D^{E_{1}},D^{R_{1,2}},Q).P( D^{E_{1}}|D^{R_{1,2}},Q).P(D^{R_{1,2}}|Q).P(Q)}{P(Q)}\\
    & \stackrel{\text{$rank$}}{=}   P(D^{E_2}| D^{R_{1,2}},Q).P( D^{E_{1}}|D^{R_{1,2}},Q).P(D^{R_{1,2}}|Q)\\
    & \stackrel{\text{$rank$}}{\propto} P(D^{E_2}| D^{R_{1,2}},Q^{E_2}).P(D^{E_1}|D^{R_{1,2}},Q^{E_1}).P(D^{R_{1,2}}|Q^{R_{1,2}})\\
\end{split}\\
\end{align}

We consider conditional independence between entity representations within the joint posterior, i.e., the probability of a given entity representation $D^{E_i}$ being relevant given a E-R query is independent of knowing that entity $D^{E_{i+1}}$ is relevant as well. As an example, consider the query ``action movies starring a British actor''. Retrieving entity representations for ``action movies'' is independent of knowing that <Tom Hardy> is relevant to the sub-query ``British actor''. However, it is not independent of knowing the set of relevant relationships for sub-query ``starring''. If a given action movie is not in the set of relevant entity-pairs for ``starring'' it does not make sense to consider it as relevant. Consequently, $P(D^{E_2}| D^{E_{1}},D^{R_{1,2}},Q) = P(D^{E_2}| D^{R_{1,2}},Q)$.

Since E-R queries can be decomposed in constituent entity and relationship sub-queries, ranking candidate tuples using the joint posterior $P(D^{E_2}, D^{E_{1}},D^{R_{1,2}}|Q)$ is rank proportional to the product of conditional probabilities on the corresponding entity and relationship sub-queries $Q^{E_2}$, $Q^{E_1}$ and $Q^{R_{1,2}}$.

We now consider a longer E-R query aiming to retrieve a triple of connected entities. This query has three entity sub-queries and two relationship sub-queries, thus $|Q|=5$. As we previously explained, when there are more than one relationship sub-queries we need to join entity-pairs relevant to each relationship sub-query that have one entity in common. From a probabilistic point of view this can be seen as conditional dependence from the entity-pairs retrieved from the previous relationship sub-query, i.e. $P(D^{R_{2,3}}|D^{R_{1,2}},Q) \neq P(D^{R_{2,3}}|Q)$.  To rank entity and relationship candidates we need to calculate the following joint posterior:

\begin{align}
\begin{split}\label{erprob}
   P(D^{E_3},  D^{E_{2}},&D^{E_{1}}, D^{R_{2,3}},D^{R_{1,2}}|Q)\stackrel{\text{$rank$}}{=} P(D^{E_3}| D^{E_{2}},D^{E_{1}},D^{R_{2,3}},D^{R_{1,2}},Q).\\
   & P( D^{E_{3}}|D^{E_{2}},D^{R_{2,3}},D^{R_{1,2}},Q).P(D^{E_{1}}|D^{R_{2,3}},D^{R_{1,2}},Q). \\
   & P(D^{R_{2,3}}|D^{R_{1,2}},Q).P(D^{R_{1,2}}|Q)\\
&  \ \ \ \ \ \ \ \ \ \ \ \ \ \ \ \ \ \ \ \ \ \ \ \ \ \  \stackrel{\text{$rank$}}{=}   P(D^{E_3}| D^{R_{2,3}},Q).P( D^{E_{2}}|D^{R_{2,3}},D^{R_{1,2}},Q).\\
& P(D^{E_{1}}|D^{R_{1,2}},Q).P(D^{R_{2,3}}|D^{R_{1,2}},Q).P(D^{R_{1,2}}|Q)\\
&  \ \ \ \ \ \ \ \ \ \ \ \ \ \ \ \ \ \ \ \ \ \ \ \ \ \  \stackrel{\text{$rank$}}{\propto}   P(D^{E_3}| D^{R_{2,3}},Q^{E_3}).P( D^{E_{2}}|D^{R_{2,3}},D^{R_{1,2}},Q^{E_{2}}).\\
& P(D^{E_{1}}|D^{R_{1,2}},Q^{E_{1}}).P(D^{R_{2,3}}|D^{R_{1,2}},Q^{R_{2,3}}).P(D^{R_{1,2}}|Q^{R_{1,2}})\\
\end{split}\\
\end{align}
  
When compared to the previous example, the joint posterior for $|Q|=5$  shows that entity candidates for  $D^{E_{2}}$ are conditional dependent of both $D^{R_{2,3}}$ and $D^{R_{1,2}}$. In other words, entity candidates for $D^{E_{2}}$ must belong to entity-pairs candidates for both relationships representations that are connected with $E_2$, i.e. $D^{R_{2,3}}$ and $D^{R_{1,2}}$. 

We are now able to make a generalization of E-R retrieval as a factorization of conditional probabilities of a joint probability of entity representations $D^{E_i}$, relationship representations $D^{R_{i-1,i}}$, entity sub-queries $Q^{E_{i}}$ and relationship sub-queries $Q^{R_{i-1,i}}$. These set of random variables and their conditional dependencies can be easily represented in a probabilistic directed acyclic graph,i.e. a Bayesian network \cite{pearl1985bayesian}. 

In Bayesian networks, nodes represent random variables while edges represent conditional dependencies. Every other nodes that point to a given node are considered parents. Bayesian networks define the joint probability of a set of random variables as a factorization of the conditional probability of each random variable conditioned on its parents. Formally, $P(X_{1},\ldots ,X_{n})=\prod _{i=1}^{n}P(X_{i}|pa_{i})$, where $pa_{i}$ represents all parent nodes of $X_i$.

\begin{figure}
\centering
\begin{subfigure}{.4\linewidth}
    \centering
    \includegraphics[width=.8\textwidth]{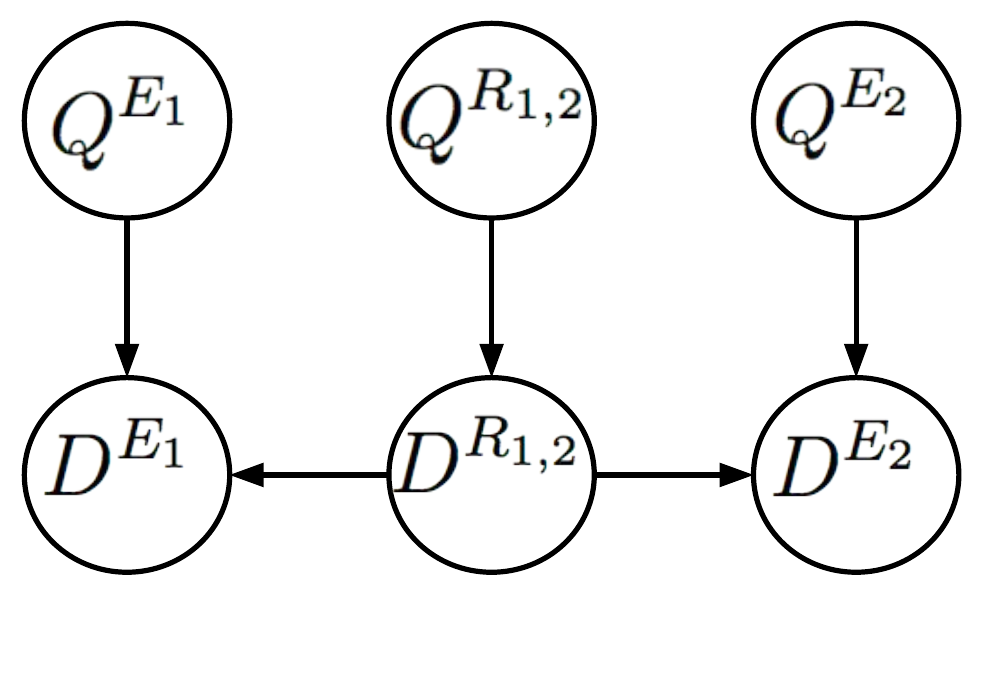}
    \caption{$|Q|=3$}
\end{subfigure}
    \hfill
\begin{subfigure}{.55\linewidth}
    \centering
	\includegraphics[width=1.0\textwidth]{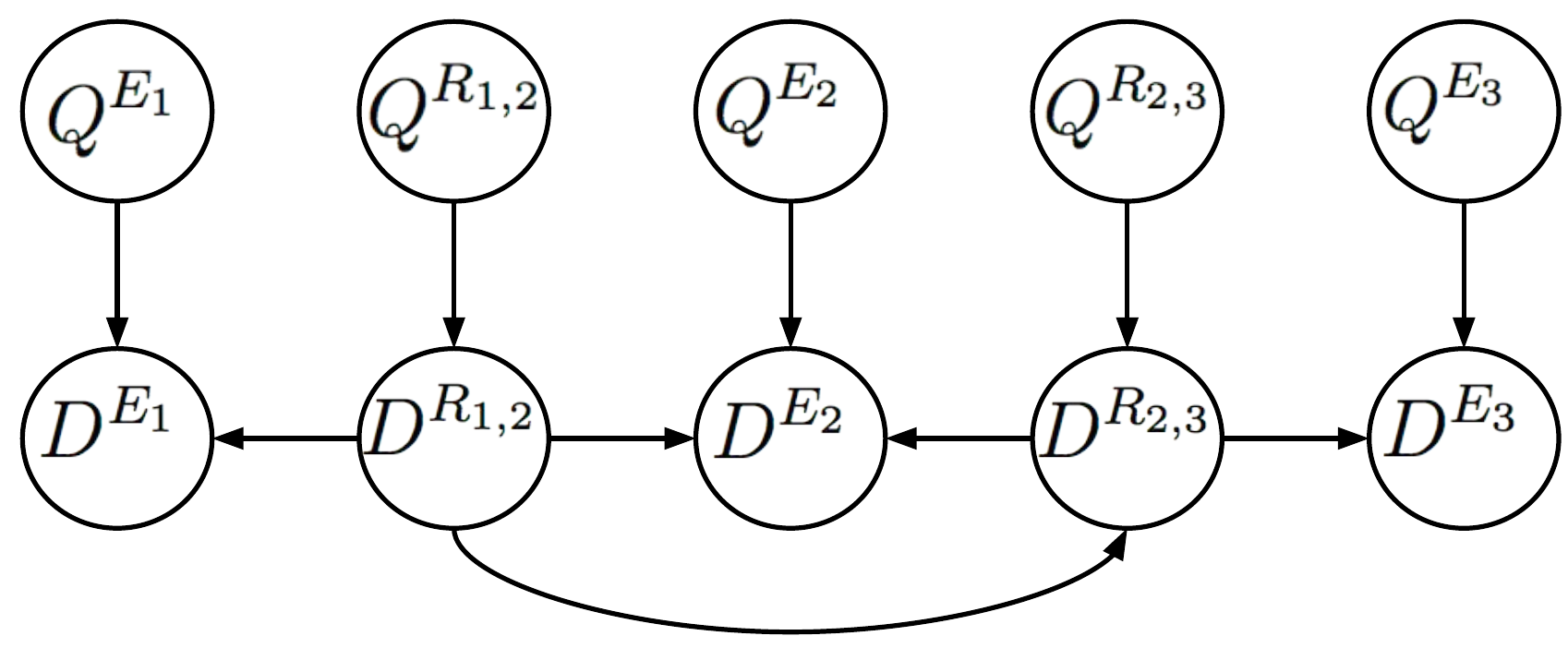}
    \caption{$|Q|=5$}
\end{subfigure}
   \hfill
   \bigskip
\begin{subfigure}{1.0\linewidth}
  \centering
	\includegraphics[width=0.8\textwidth]{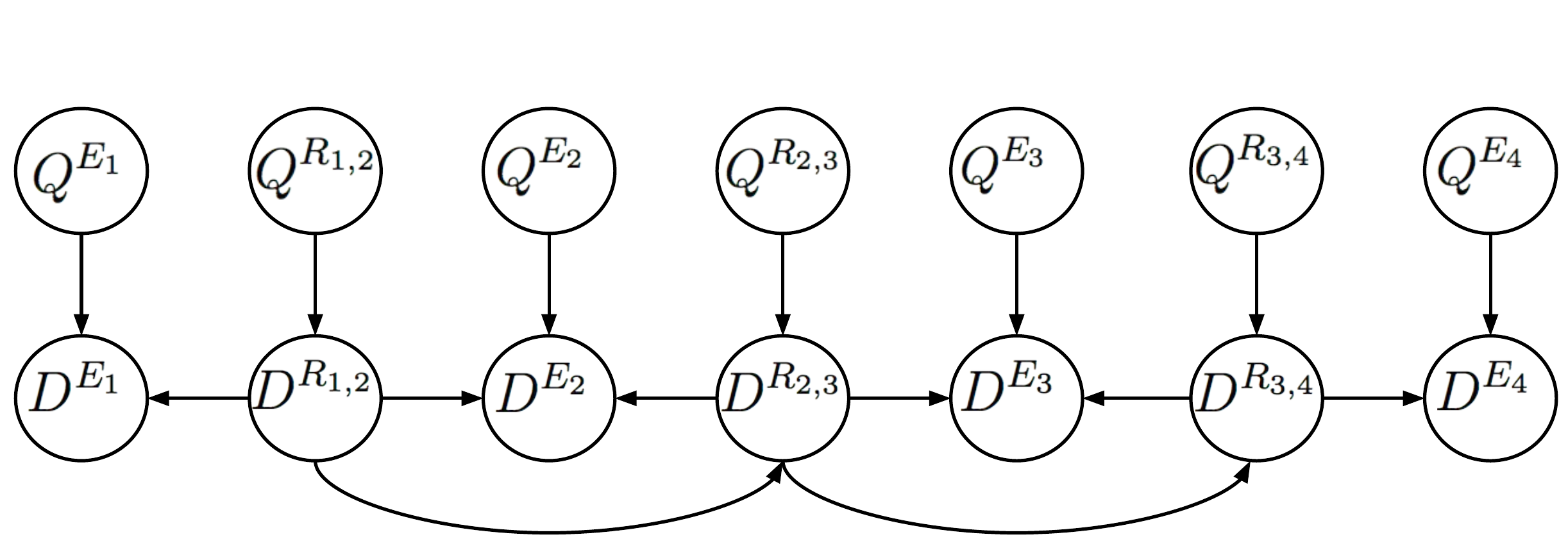}
  \caption{$|Q|=7$}
\end{subfigure} 
\caption{Bayesian networks for E-R Retrieval with queries of different lengths.}
\label{fig:bayesian_er}
\end{figure}

Figure \ref{fig:bayesian_er} depicts the representation of E-R retrieval for different query lengths $|Q|$ using Bayesian networks. We easily conclude that graphical representation contributes to establish a few guidelines for modeling E-R retrieval. First, each sub-query points to the respective document node. Second, relationship document nodes always point to the contiguous entity representations. Last, when there are more than one relationship sub-query, relationship documents also point to the subsequent relationship document.

Once we draw the graph structure for the number of sub-queries in $Q$ we are able to compute a product of conditional probabilities of each node given its parents. Adapting the general joint probability formulation of Bayesian networks to E-R retrieval we come up with the following generalization:

\begin{equation}
P(D^{E},D^{R}|Q) \stackrel{\text{$rank$}}{=}  \prod_{i=1}^{\frac{|Q|+1}{2}} P(D^{E_i}|D^{R_{i-1,i}},D^{R_{i,i+1}}, Q^{E_i})  \prod_{i=1}^{\frac{|Q|-1}{2}}   P(D^{R_{i,i+1}}|D^{R_{i-1,i}},   Q^{R_{i,i+1}})  
\end{equation}

We denote $D^{R}$ as the set of all candidate relationship documents in the graph and $D^{E}$ the set of all candidate entity documents in the graph. In Information Retrieval is often convenient to work in the log-space as it does not affect ranking and transforms the product of conditional probabilities in a summation, as follows:

\begin{align}
\begin{split}\label{eq:erlog}
P(D^{E},D^{R}|Q)   & \stackrel{\text{$rank$}}{=}  \text{log} \ P(D^{E},D^{R}|Q) \\
& \stackrel{\text{$rank$}}{=}  \sum_{i=1}^{\frac{|Q|+1}{2}}  \text{log} P(D^{E_i}|D^{R_{i-1,i}},D^{R_{i,i+1}}, Q^{E_i}) + \sum_{i=1}^{\frac{|Q|-1}{2}}    \text{log} P(D^{R_{i,i+1}}|D^{R_{i-1,i}},   Q^{R_{i,i+1}}) \\
\end{split}\\
\end{align}

In essence, E-R retrieval is an extension, or a more complex case, of object-retrieval where besides ranking objects we need to rank tuples of objects that satisfy the relationship expressed in the E-R query. This requires creating representations of both entities and relationships by fusing information spread across multiple raw documents.  We hypothesize that it should be possible to generalize the term dependence models to represent entity-relationships and achieve effective E-R retrieval without entity or relationship type restrictions (e.g. categories) as it happens with the Semantic Web based approaches.

\section{Early Fusion}
Traditional ad-hoc document retrieval approaches create direct term-based representations of raw documents. A retrieval model (e.g. Language Models) is then used to match the information need, expressed as a keyword query, against those representations. However, E-R retrieval requires collecting evidence for both entities and relationships that can be spread across multiple documents. It is not possible to create direct term-based representations. Raw documents serve as proxy to connect queries with entities and relationships.

As described in previous section, E-R queries can be formulated as a sequence of multiple entity queries $Q^{E}$ and relationship queries $Q^{R}$. In a Early Fusion approach, each of these queries should match against a previously created term-based representation. Since there are two types of queries, we propose to create two types of term-based representations, one for entities and other for relationships. It can be thought as creating two types of meta-documents $D^{E}$ and $D^{R}$. A meta-document $D^{E_i}$ is created by aggregating the context terms of the occurrences of $E_i$ across the raw document collection. On the other hand, for each each pair of entities $E_{i-1}$ and $E_i$  that co-occur close together across the raw document collection we aggregate context terms that describe the relationship to create a meta-document $D^{R_{i-1,i}}$. 

Our Early Fusion design pattern for E-R retrieval is inspired in \textit{Model 1} of \cite{balog2006formal} for single entity retrieval.
 In our approach we focus on sentence level information about entities and relationships although the design pattern can be applied to more complex segmentations of text (e.g. dependency parsing). We rely on Entity Linking methods for disambiguating and assigning unique identifiers to entity mentions on raw documents $D$. We collect entity contexts across the raw document collection and index them in the \textit{entity index}. The same is done by collecting and indexing entity pair contexts in the \textit{relationship index}.  

We define the (pseudo) frequency of a term $t$ for an entity meta-document $D^{E_i}$ as follows:

\begin{equation} \label{tf_e}
f(t,D^{E_i}) =  \sum_{j=1}^{n} f(t,E_i,D_j) w(E_i,D_j)
\end{equation}

where $n$ is the total number of raw documents in the collection,  $f(t,E_i,D_j)$ is the term frequency in the context of the entity $E_i$ in a raw document $D_j$.  $w(E_i,D_j)$ is the entity-document association weight that corresponds to the weight of the document $D_j$ in the mentions of the entity $E_i$ across the raw document collection. Similarly, the term (pseudo) frequency of a term $t$ for a relationship meta-document $D^{R_{i-1,i}}$ is defined as follows:

\begin{equation} \label{tf_r}
f(t,D^{R_{i-1,i}}) =  \sum_{j=1}^{n} f(t,R_{i-1,i},D_j) w(R_{i-1,i},D_j)
\end{equation}

where $f(t,R_{i-1,i},D_j$ is the term frequency in the context of the pair of entity mentions corresponding to the relationship $R_{i-1,i}$ in a raw document $D_j$ and $w(R_{i-1,i},D_j)$ is the relationship-document association weight. In this work we use binary associations weights indicating the presence/absence of an entity mention in a raw document, as well as for a relationship. However, other weight methods can be used.

The relevance score for an entity tuple $T_E$ can then be calculated using the posterior $P(D^{E},D^{R}|Q)$ defined in previous section (equation \ref{eq:erlog}). We calculate the individual conditional probabilities  as a product of a retrieval score with an association weight. Formally we consider:

\begin{align}
\begin{split}\label{ef_eq}
 & \text{log}P(D^{E_i}|D^{R_{i-1,i}},D^{R_{i,i+1}}, Q^{E_i}) = score(D^{E_i}, Q^{E_i}) w(E_i, R_{i-1,i}, R_{i,i+1}) \\
	& \text{log} P(D^{R_{i,i+1}}|D^{R_{i-1,i}}, Q^{R_{i,i+1}}) = score(D^{R_{i,i+1}}, Q^{R_{i,i+1}}) w(R_{i,i+1}, R_{i-1,i}) \\
\end{split}\\
\end{align}

where $score(D^{R_{i,i+1}}, Q^{R_{i,i+1}})$ represents the retrieval score resulting of the match of the query terms of a relationship sub-query $Q^{R_{i,i+1}}$ and a relationship meta-document $D^{R_{i,i+1}}$. The same applies to the retrieval score $score(D^{E_i}, Q^{E_i})$ which corresponds to the result of the match of an entity sub-query $Q^{E_i}$ with a entity meta-document $D^{E_i}$. For computing both $score(D^{R_{i,i+1}}, Q^{R_{i,i+1}})$ and $score(D^{E_i}, Q^{E_i})$ any retrieval model can be used. 

We use a binary association weight for $w(E_i, R_{i-1,i}, R_{i,i+1})$ which represents the presence of a relevant entity $E_i$ to a sub-query $Q^{E_i}$ in its contiguous relationships in the Bayesian network, i.e.  $R_{i-1,i}$ and $R_{i,i+1}$ which must be relevant to the sub-queries $Q^{R_{i-1,i}}$ and $Q^{R_{i,i+1}}$. This entity-relationship association weight is the building block that guarantees that two entities relevant to sub-queries $Q^E$ that are also part of a relationship relevant to a sub-query $Q^R$ will be ranked higher than tuples where just one or none of the entities are relevant to the entity sub-queries $Q^E$. On the other hand, the entity-relationship association weight $w(R_{i,i+1}, R_{i-1,i})$ guarantees that consecutive relationships share one entity between them in order to create triples or 4-tuples of entities for longer E-R queries ($|Q|>3$).

The relevance score of an entity tuple $T_E$ given a query $Q$ is calculated by summing individual relationship and entity relevance scores for each $Q^{R_{i-1,i}}$ and $Q^{E_i}$ in $Q$. We define the score for a tuple $T_E$ given a query $Q$ as follows:

\begin{align}
\begin{split}\label{efd_eq}
  P(D^{E},D^{R}|Q) \stackrel{\text{$rank$}}{=} & \sum_{i=1}^{\frac{|Q|+1}{2}} score(D^{E_i}, Q^{E_i}) w(E_i, R_{i-1,i}, R_{i,i+1})+\\
	& \sum_{i=1}^{\frac{|Q|-1}{2}}  score(D^{R_{i,i+1}}, Q^{R_{i,i+1}}) w(R_{i,i+1}, R_{i-1,i})\\
\end{split}\\
\end{align}

Considering Dirichlet smoothing unigram Language Models (LM) the constituent retrieval scores can be computed as follows:
\begin{equation}
score_{LM}(D^{R_{i,i+1}}, Q^{R_{i,i+1}}) = \sum_{t \in D^{R_{i,i+1}} \cap Q^{R_{i,i+1}} } \text{log}  \left (\frac{ f(t,D^{R_{i,i+1}}) +  \frac{f(t,C^R)}{|C^R|}\mu^R}{|D^{R_{i,i+1}}| + \mu^R} \right ) 
\end{equation}

\begin{equation}
score_{LM}(D^{E_i}, Q^{E_i}) = \sum_{t \in D^{E_i}  \cap Q^{E_i}} \text{log} \left (\frac{f(t,D^{E_i})
 + \frac{ f(t,C^E)}{|C^E|}\mu^E }{|D^{E_i}| + \mu^E} \right )
\end{equation}

where $t$ is a term of a sub-query $Q^{E_i}$ or $Q^{R_{i,i+1}}$, $f(t,D^{E_i})$ and
$f(t,D^{R_{i,i+1}})$ are the (pseudo) frequencies defined in equations \ref{tf_e} and \ref{tf_r}. The collection frequencies $f(t,C^E)$, $f(t,C^R)$ represent the frequency of the term $t$ in either the \textit{entity index} $C^E$ or in the \textit{relationship index} $C^R$. $|D^{E_i}|$ and$|D^{R_{i,i+1}}|$ represent the total number of terms in a meta-document while $|C^R|$ and $|C^E|$ represent the total number of terms in a collection of meta-documents. Finally, $\mu^E$ and $\mu^R$ are the Dirichlet prior for smoothing which generally corresponds to the average document length in a collection.

\subsection{Association Weights}

In Early Fusion there are three association weights: $w(R_{i,i+1},D_j)$, $w(E_i,D_j)$ and $w(E_i, R_{i,i+1})$. The first two represent document associations which determine the weight a given raw document contributes to the relevance score of a particular entity tuple $T_E$. The last one is the entity-relationship association which indicates the strength of the connection of a given entity $E_i$ within a relationship $R_{i,i+1}$. 

In our work we only consider binary association weights but other methods could be used. According to the binary method we define the weights as follows:

\begin{equation}
w(R_{i,i+1},D_j) = 1 \ \text{if}\ R(E_i,E_{i+1})\ \in\ D_j \ , 0 \ \text{otherwise}
\end{equation}
\begin{equation}
w(E_i,D_j) = 1 \ \text{if}\ E_i\ \in\ D_j \ , 0 \ \text{otherwise}
\end{equation}
\begin{equation}
w(E_i, R_{i-1,i},R_{i,i+1}) = 1 \ \text{if}\ E_i\ \in\ D^{R_{i-1,i}} and \  E_i\ \in\ D^{R_{i,i+1}} \ , 0 \ \text{otherwise}
\end{equation}
\begin{equation}
w(R_{i,i+1},R_{i-1,i}) = 1 \ \text{if}\ E_i\ \in\ D^{R_{i-1,i}} and \  E_i\ \in\ D^{R_{i,i+1}} \ , 0 \ \text{otherwise}
\end{equation}

 Under this approach the weight of a given association is independent of the number of times an entity or a relationship occurs in a document. A more general approach would be to assign real numbers to the association weights depending on the strength of the association \cite{balog2012expertise}. For instance, uniform weighting would be proportional to the inverse of the number of documents where a given entity or relationship occurs. Other option could be a TF-IDF approach.

\subsection{Discussion}
This approach is flexible enough to allow using any retrieval method to compute individual retrieval scores between document and query nodes in a E-R graph structure. When using Language Models (LM) or BM25 as scoring functions, these design patterns can be used to create unsupervised baseline methods for E-R retrieval (e.g. EF-LM, EF-BM25, LF-LM, LF-BM25, etc.).

There is some overhead over traditional document search, since we need to create two E-R dedicated indexes that will store entity and relationship meta-documents. The \textit{entity} index is created by harvesting the context terms in the proximity of every occurrence of a given entity across the raw document collection. This process must be carried for every entity in the raw document collection. A similar process is applied to create the relationship index. For every two entities occurring close together in a raw document we extract the text between both occurrences as a term-based representation of the relationship between the two. Once again, this process must be carried for every pair of co-occurring entities in sentences across the raw document collection.  

One advantage of Early Fusing lies in its flexibility as we need to create two separate indexes for E-R retrieval it is possible to combine data from multiple sources in seamless way. For instance, one could use a well established knowledge base (e.g. DBpedia) as entity index and use a specific collection, such as a news collection or a social media stream, for harvesting relationships having a more transient nature.

The challenge inherent to the problem of E-R retrieval is the size of the search space. Although the E-R problem is formulated as a sequence of independent sub-queries, the results of those sub-queries must be joined together. Consequently, we have a multi-dimensional search space in which we need to join results based on shared entities. 

This problem becomes particularly hard when sub-queries are short and contain very popular terms. Let us consider ``actor'' as $Q^{E_i}$, there will be many results to this sub-query, probably thousands. There is a high probability that will need to process thousands of sub-results before finding one entity that is also relevant to the relationship sub-query $Q^{R_{i-1,i}}$. If at the same time we have computational power constraints, we will probably apply a strategy of just considering top k results for each sub-query which can lead to reduced recall in the case of short sub-queries with popular terms.

\section{Entity-Relationship Dependence Model}

In this section we present the Entity-Relationship Dependence Model (ERDM), a novel supervised Early Fusion-based model for E-R retrieval. Recent approaches to entity retrieval \cite{zhiltsov2015fielded,nikolaev2016parameterized,hasibi2016exploiting} have demonstrated that using models based on Markov Random Field (MRF) framework for retrieval \cite{metzler2005markov} to incorporate term dependencies can improve entity search performance. This suggests that MRF could be used to model E-R query term dependencies among entities and relationships documents.

One of the advantages of the MRF framework for retrieval is its flexibility, as we only need to construct a graph $G$ representing dependencies to model, define a set of non-negative potential functions $\psi$ over the cliques of $G$ and to learn the parameter vector $\Lambda$ to score each document $D$ by its unique and unnormalized joint probability with $Q$ under the MRF \cite{metzler2005markov}.

The non-negative potential functions are defined using an exponential form $\psi(c;\Lambda) = \text{exp} [\lambda_c f(c)]$, where $\lambda_c$  is a feature weight, which is a free parameter in the model, associated with feature function $f(c)$. Learning to rank is then used to learn the feature weights that minimize the loss function. The model allows parameter and feature functions sharing across cliques of the same configuration, i.e. same size and type of nodes (e.g. 2-cliques of one query term node and one document node).

\subsection{Graph Structures}

The Entity-Relationship Dependence Model (ERDM) creates a MRF for modeling implicit dependencies between sub-query terms, entities and relationships. Each entity and each relationship are modeled as document nodes within the graph and edges reflect term dependencies. Contrary to traditional ad-hoc retrieval using MRF (e.g. SDM), where the objective is to compute the posterior of a single document given a query, the ERDM allows the computation of a joint posterior of multiple documents (entities and relationships) given a E-R query which consists also of multiple sub-queries.

\begin{figure}[h] 
\centering
\includegraphics[width=0.6\textwidth]{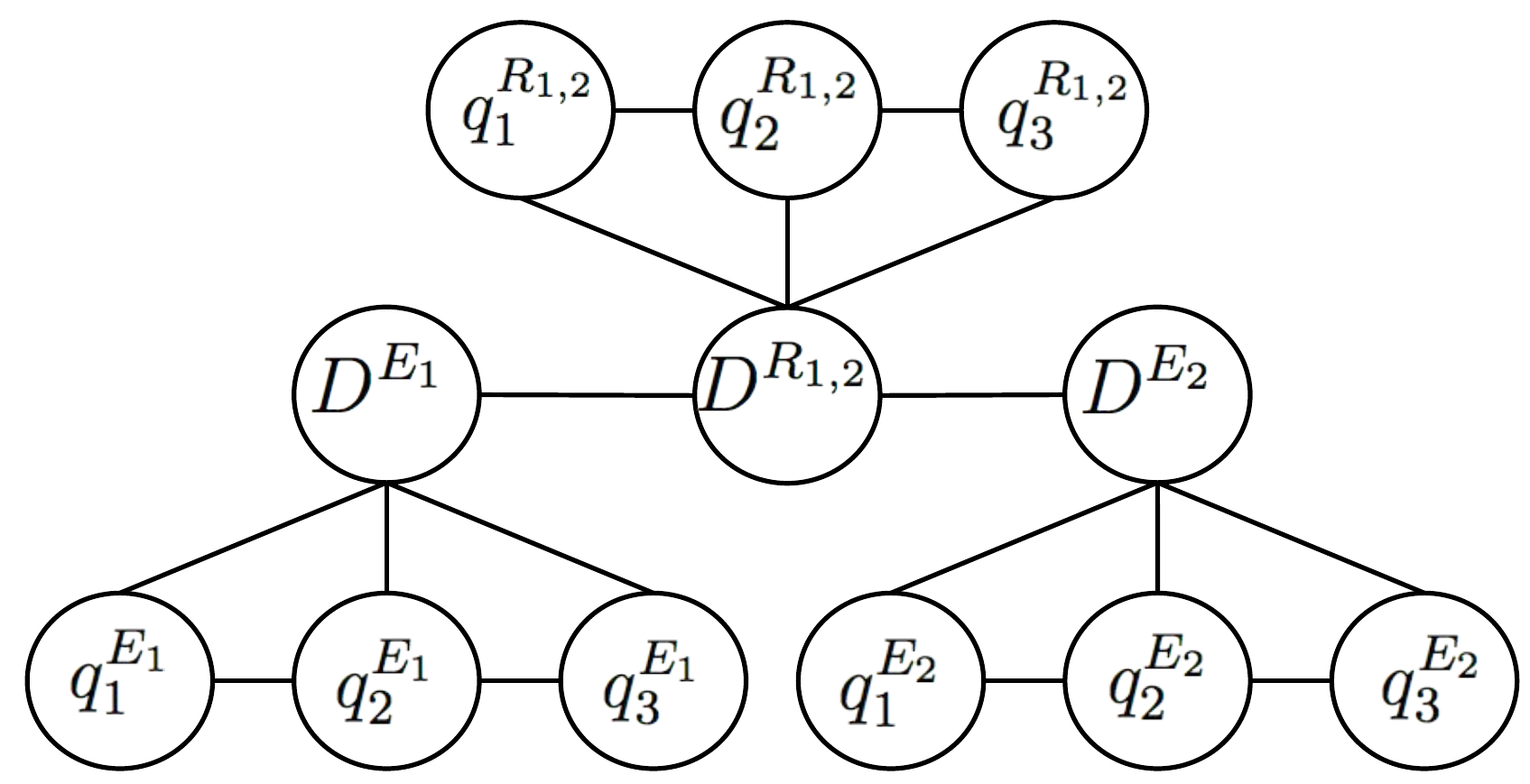}
\caption{Markov Random Field dependencies for E-R retrieval, $|Q|=3$.}\label{fig:erdm3}
\end{figure}

\begin{figure}[h] 
\centering
\includegraphics[width=0.9\textwidth]{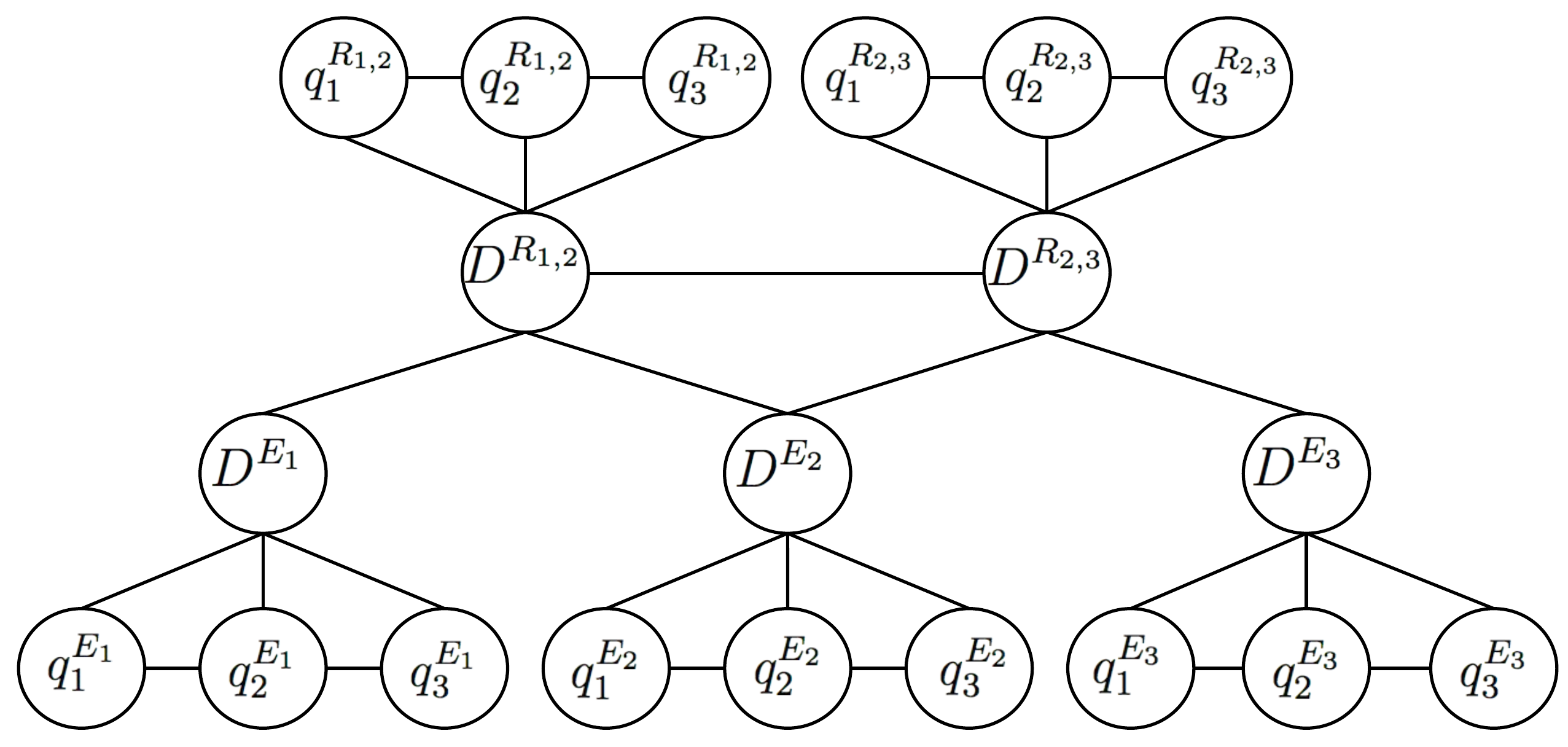}
\caption{Markov Random Field dependencies for E-R retrieval, $|Q|=5$.}\label{fig:erdm5}
\end{figure}

The graph structures of the ERDM for two E-R queries, one with $|Q|=3$ and other with $|Q|=3$ are depicted in Figure \ref{fig:erdm3} and Figure \ref{fig:erdm5}, respectively. Both graph structures contain two different types of query nodes and document nodes: entity query and relationship query nodes, $Q^E$ and $Q^R$, plus entity and relationship document nodes, $D^E$ and $D^R$. Within the MRF framework, $D^E$ and $D^R$ are considered ``documents'' but they are not actual real documents but rather objects representing an entity or a relationship between two entities. Unlike real documents, these objects do not have direct and explicit term-based representations. Usually, it is necessary to gather evidence across multiple real documents that mention the given object, in order to be able to match
them against keyword queries. Therefore, ERDM can be seen as Early Fusion-based retrieval model. The existence of two different types of documents implies two different indexes: the \textit{entity} index and the \textit{relationship} index.

The relationship-specific dependencies of ERDM are found in the 2-cliques formed by one entity document and one relationship document: $D^{E_{i-1}}$ - $D^{R_{i-1,i}}$, $D^{E_i}$- $D^{R_{i-1,i}}$  and for $|Q|=5$,  $D^{E_i}$- $D^{R_{i,i+1}}$ and $D^{R_{i-1,i}}$ - $D^{R_{i,i+1}}$. The graph structure does not need to assume any explicit dependence between entity documents given a relationship document. They have an implicit connection through the dependencies with the relationship document. The likelihood of observing an entity document $D^{E_i}$ given a relationship document $D^{R_{i-1,i}}$ is not affected by the observation of any other entity document. 

Explicit dependence between the two entity documents could be used to represent the direction of the relationship between the two entities. To support this dependence, relationship documents would need to account the following constraint: $R(E_{i-1},E_i) \neq R(E_{i},E_{i-1}), \text{ } \forall \text{ }  D^{R_{i-1,i}} \in C^R$, with $C^R$ representing the \textit{relationship} index. Then, we would compute an ordered feature function between entities in a relationship, similar to the ordered bigram feature function in SDM. In this work, we do not explicitly model asymmetric relationships. For instance, if a user searches for the relationship entity A ``criticized'' entity B but was in fact entity B who criticized entity A we assume that the entity tuple   <entity A, entity B> is still relevant for the information need expressed in the E-R query.

ERDM follows the SDM \cite{metzler2005markov} dependencies between query terms and documents due to its proved effectiveness in multiple contexts. Therefore, ERDM assumes a dependence between neighboring sub-query terms: 

\begin{equation}
P(q^{E_i}_{j}|D^{E_i},q^{E_i}_{j \neq l}) = P(q^{E_i}_{j}|D^{E_i},q^{E_i}_{j-1},q^{E_i}_{j+1})
\end{equation}

\begin{equation}
P(q^{R_{i-1,i}}_{j}|D^{R_{i-1,i}},q^{R_{i-1,i}}_{j \neq l},D^{E_i}) = P(q^{R_{i-1,i}}_{j}|D^{R_{i-1,i}},q^{R_{i-1,i}}_{j-1},q^{R_{i-1,i}}_{j+1})
\end{equation}

MRF for retrieval requires the definition of the sets of cliques (maximal or non-maximal) within the graph that one or more feature functions is to be applied to. The set of cliques in ERDM containing at least one document are the following:

\begin{itemize}
\item $T^{E}$ - set of 2-cliques containing an entity document node and exactly one term in a entity sub-query.

\item $O^{E}$ - set of 3-cliques containing an entity document node and two ordered terms in a entity sub-query.

\item $T^{R}$ - set of 2-cliques containing a relationship document node and exactly one term in a relationship sub-query.

\item $O^{R}$ - set of 3-cliques containing a relationship document node and two ordered terms in a relationship sub-query.

\item $S^{ER}$ - set of 2-cliques containing one entity document node and one relationship document node.

\item $S^{RER}$ - set of 3-cliques containing one entity document node and two consecutive relationship document nodes.

\end{itemize}

The joint probability mass function of the MRF is computed using the set of potential functions over the configurations of the maximal cliques in the graph \cite{metzler2005markov}. Non-negative potential functions are constructed from one or more real valued feature functions associated with the respective feature weights using an exponential form.

\subsection{Feature Functions}

ERDM has two types of feature functions: textual and non-textual. Textual feature functions measure the textual similarity between one or more sub-query terms and a document node. Non-textual feature functions measure compatibility between entity and relationship documents, i.e., if they share a given entity.

\begin{table}[h]
\centering
\caption{Clique sets and associated feature functions by type and input nodes.}
\label{tab:cliqfeats}
\begin{tabular}{|l|l|l|l|}
\hline
Clique Set & Feature Functions  & Type & Input Nodes \\ \hline
$T^{E}$   & $f^{E}_T$  & Textual &  $\{q^{E_i}_j,D^{E_i}\}$           \\ \hline
$O^{E}$ & $f^{E}_O$ and $f^{E}_U$ &Textual &   $\{q^{E_i}_j,q^{E_i}_{j+1},D^{E_i}\}$           \\ \hline
  $T^{R}$  & $f^{R}_T$  &Textual &   $\{q^{R_{i-1,i}}_j,D^{R_{i-1,i}}\}$           \\ \hline

$O^{R}$& $f^{R}_O$ and $f^{E}_U$   &Textual &   $\{q^{R_{i-1,i}}_j,q^{R_{i-1,i}}_{j+1} ,D^{R_{i-1,i}}\}$           \\ \hline
$S^{ER}$   & $f^{ER}_S$  &Non-textual &   $\{D^{E_i},D^{R_{i-1,i}}\}$           \\ \hline
$S^{RER}$   & $f^{RER}_S$  &Non-textual &   $\{D^{E_i},D^{R_{i-1,i}},D^{R_{i,i+1}}\}$           \\ \hline

\end{tabular}
\end{table}

Table \ref{tab:cliqfeats} presents an overview of the feature functions associated with clique sets and the type of input nodes. Although we could define a wide set of different feature functions, we decided to adapt SDM textual feature functions to ERDM clique configurations. Therefore we define unigram based feature functions $f^{E}_T$  and $f^{R}_T$ to 2-cliques containing a single sub-query term and a entity or relationship document node. 

For 3-cliques containing consecutive sub-query terms and a document node, we define two feature functions. One considers consecutive sub-query terms and matches ordered bigrams with entity or relationship documents. This feature function is denoted as $f^{E}_O$ and $f^{R}_O$, depending if the clique is $O^{E}$ or $O^{R}$. The second feature function matches bigrams with documents using an unordered window of 8 terms ($uw8$), i.e., it matches bigrams with documents if the two terms of the bigram occur with a maximum of 6 other terms between each other. This feature function is denoted as $f^{E}_U$ and $f^{R}_U$, depending if the clique is $O^{E}$ or $O^{R}$.

For each textual feature function we decided to use two variants: Dirichlet smoothing Language Models (LM) and BM25. We now present the summary of the textual feature functions used in this work.

\ \\
\textbf{LM}-T-E\\
\resizebox{0.5 \linewidth}{!} 
{
$f^{E}_{T,LM}(q^{E_i}_j,D^{E_i}) =  \text{log} \left (\frac{f(q^{E_i}_j,D^{E_i})
 + \frac{ f(q^{E_i}_j,C^E)}{|C^E|}\mu^E }{|D^{E_i}| + \mu^E} \right )$
 }\\

\ \\ \textbf{LM}-O-E\\

\resizebox{0.7 \linewidth}{!} 
{
$f^{E}_{O,LM}(q^{E_i}_j,q^{E_i}_{j+1},D^{E_i}) = \text{log} \left (\frac{f_{\#1}(q^{E_i}_j,q^{E_i}_{j+1},D^{E_i})
 + \frac{ f_{\#1}(q^{E_i}_j,q^{E_i}_{j+1},C^E)}{|C^E|}\mu^E }{|D^{E_i}| + \mu^E} \right )$
 }
 
\ \\ \textbf{LM}-U-E\\

\resizebox{0.7 \linewidth}{!} 
{
$f^{E}_{U,LM}(q^{E_i}_j,q^{E_i}_{j+1},D^{E_i}) =  \text{log} \left (\frac{f_{\#uw8}(q^{E_i}_j,q^{E_i}_{j+1},D^{E_i})
 + \frac{ f_{\#uw8}(q^{E_i}_j,q^{E_i}_{j+1},C^E)}{|C^E|}\mu^E }{|D^{E_i}| + \mu^E} \right )$
 }

\ \\ \textbf{LM}-T-R\\

\resizebox{0.7 \linewidth}{!} 
{
$f^{R}_{T,LM}(q^{R_{i-1,i}}_j,D^{R_{i-1,i}}) = \text{log}  \left (\frac{ f(q^{R_{i-1,i}}_j,D^{R_{i-1,i}}) +  \frac{f(q^{R_{i-1,i}}_j,C^R)}{|C^R|}\mu^R}{|D^{R_{i-1,i}}| + \mu^R} \right ) $
 }

\ \\ \textbf{LM}-O-R\\

\resizebox{1.0 \linewidth}{!} 
{
$f^{R}_{O,LM}(q^{R_{i-1,i}}_j,q^{R_{i-1,i}}_{j+1} ,D^{R_{i-1,i}}) = \text{log}  \left (\frac{ f_{\#1}(q^{R_{i-1,i}}_j,q^{R_{i-1,i}}_{j+1},D^{R_{i-1,i}}) +  \frac{f_{\#1}(q^{R_{i-1,i}}_j,q^{R_{i-1,i}}_{j+1},C^R)}{|C^R|}\mu^R}{|D^{R_{i-1,i}}| + \mu^R} \right ) $
 }
 
\ \\ \textbf{LM}-U-R\\

\resizebox{1.0 \linewidth}{!} 
{
$f^{R}_{U,LM}(q^{R_{i-1,i}}_j,q^{R_{i-1,i}}_{j+1} ,D^{R_{i-1,i}}) = \text{log}  \left (\frac{ f_{\#uw8}(q^{R_{i-1,i}}_j,q^{R_{i-1,i}}_{j+1},D^{R_{i-1,i}}) +  \frac{f_{\#uw8}(q^{R_{i-1,i}}_j,q^{R_{i-1,i}}_{j+1},C^R)}{|C^R|}\mu^R}{|D^{R_{i-1,i}}| + \mu^R} \right ) $
 }

\ \\
\ \\
Here, $f(q^{E_i}_j,D^{E_i})$ and $f(q^{R_{i-1,i}}_j,D^{R_{i-1,i}})$ represent the sub-query term frequencies in a entity document and relationship document, respectively.  The collection frequencies $f(q^{E_i}_j,C^E)$, $f(q^{R_{i-1,i}}_j,C^R)$ represent the frequency of sub-query term in either the \textit{entity} index $C^E$ or in the \textit{relationship} index $C^R$. The variants of these functions $f_{\#1}$ and $f_{\#uw8}$ represent ordered and unordered bigram matching frequency.  $|D^{E_i}|$ and$|D^{R_{i,i+1}}|$ represent the total number of terms in a meta-document while $|C^R|$ and $|C^E|$ represent the total number of terms in a collection of meta-documents. Finally, $\mu^E$ and $\mu^R$ are the Dirichlet prior for smoothing which generally corresponds to the average document length in a collection.

\ \\ \textbf{BM25}-T-E\\

\resizebox{0.7 \linewidth}{!} 
{
$f^{E}_{T,BM25}(q^{E_i}_j,D^{E_i}) =  \text{log} \frac{N^E - n(q^{E_i}_j) + 0.5}{n(q^{E_i}_j) + 0.5} \ . \   \frac{ f(q^{E_i}_j,D^{E_i}) (K_1 + 1)} {  f(q^{E_i}_j,D^{E_i}) + K_1 (1 - b + b \frac{|D^{E_i}|}{avg(|D^{E}|)})}$
 }

\ \\ \textbf{BM25}-O-E\\

\begin{align}
\begin{split}
f^{E}_{O,BM25}(q^{E_i}_j,q^{E_i}_{j+1},D^{E_i}) =&  \text{log} \frac{N^E - n_{\#1}(q^{E_i}_j,q^{E_i}_{j+1}) + 0.5}{n_{\#1}(q^{E_i}_j,q^{E_i}_{j+1}) + 0.5} \ \cdot\\
& \frac{ f_{\#1}(q^{E_i}_j,q^{E_i}_{j+1},D^{E_i}) (K_1 + 1)} {  f_{\#1}(q^{E_i}_j,q^{E_i}_{j+1},D^{E_i}) + K_1 (1 - b + b \frac{|D^{E_i}|}{avg(|D^{E}|)})}
 \end{split}\\
\end{align}

\ \\ \textbf{BM25}-U-E\\

\begin{align}
\begin{split}
f^{E}_{U,BM25}(q^{E_i}_j,D^{E_i})=&  \text{log} \frac{N^E - n_{\#uw8}(q^{E_i}_j,q^{E_i}_{j+1}) + 0.5}{n_{\#uw8}(q^{E_i}_j,q^{E_i}_{j+1}) + 0.5} \ \cdot\\
&\frac{ f_{\#uw8}(q^{E_i}_j,q^{E_i}_{j+1},D^{E_i}) (K_1 + 1)} {  f_{\#uw8}(q^{E_i}_j,q^{E_i}_{j+1},D^{E_i}) + K_1 (1 - b + b \frac{|D^{E_i}|}{avg(|D^{E}|)})}\\
\end{split}\\
\end{align}

\ \\ \textbf{BM25}-T-R\\

\begin{align}
\begin{split}
f^{R}_{T,BM25}(q^{R_{i-1,i}}_j,D^{R_{i-1,i}}) = & \text{log} \frac{N^R - n(q^{R_{i-1,i}}_j) + 0.5}{n(q^{R_{i-1,i}}_j) + 0.5} \ \cdot \\
&   \frac{ f(q^{R_{i-1,i}}_j,D^{R_{i-1,i}}) (K_1 + 1)} {  f(q^{R_{i-1,i}}_j,D^{R_{i-1,i}}) + K_1 (1 - b + b \frac{|D^{R_{i-1,i}}|}{avg(|D^{R}|)})}\\
\end{split}\\
\end{align}

\ \\ \textbf{BM25}-O-R\\

\begin{align}
\begin{split}
f^{R}_{O,BM25}(q^{R_{i-1,i}}_j,q^{R_{i-1,i}}_{j+1} ,D^{R_{i-1,i}}) = &  \text{log} \frac{N^R - n_{\#1}(q^{R_{i-1,i}}_j,q^{R_{i-1,i}}_{j+1}) + 0.5}{n_{\#1}(q^{R_{i-1,i}}_j,q^{R_{i-1,i}}_{j+1}) + 0.5} \cdot \\
& \frac{ f_{\#1}(q^{R_{i-1,i}}_j,q^{R_{i-1,i}}_{j+1} ,D^{R_{i-1,i}}) (K_1 + 1)} {  f_{\#1}(q^{R_{i-1,i}}_j,q^{R_{i-1,i}}_{j+1} ,D^{R_{i-1,i}}) + K_1 (1 - b + b \frac{|D^{R_{i-1,i}}|}{avg(|D^{R}|)})}\\
\end{split}\\
\end{align}

\ \\ \textbf{BM25}-U-R\\

\begin{align}
\begin{split}
f^{R}_{U,BM25}(q^{R_{i-1,i}}_j,q^{R_{i-1,i}}_{j+1} ,D^{R_{i-1,i}})=& \text{log} \frac{N^R - n_{\#uw8}(q^{R_{i-1,i}}_j,q^{R_{i-1,i}}_{j+1}) + 0.5}{n_{\#uw8}(q^{R_{i-1,i}}_j,q^{R_{i-1,i}}_{j+1}) + 0.5} \cdot \\
&  \frac{ f_{\#uw8}(q^{R_{i-1,i}}_j,q^{R_{i-1,i}}_{j+1} ,D^{R_{i-1,i}}) (K_1 + 1)} {  f_{\#uw8}(q^{R_{i-1,i}}_j,q^{R_{i-1,i}}_{j+1} ,D^{R_{i-1,i}}) + K_1 (1 - b + b \frac{|D^{R_{i-1,i}}|}{avg(|D^{R}|)})}\\
\end{split}\\
\end{align}

Here, $N^E$ and $N^R$ represent the total number of documents in the entity index and relationship index, respectively.  The document frequency of unigrams and bigrams is represented using $n()$,$n_{\#1}()$ and $n_{\#uw8}()$. $|D^{E_i}|$ and $|D^{R_{i-1,i}}|$ are the total number of terms in a entity or relationship document while $avg(|D^{E}|)$ and  $avg(|D^{R}|)$ are the average entity or relationship document length. $K_1$ and $b$ are free parameters usually chosen as 1.2 and 0.75, in the absence of specific optimization.

We define two non-textual features in ERDM. The first one, $f^{ER}_T$ is assigned to 2-cliques composed by one entity document and one relationship document and it is inspired in the feature function $f_E$ of Hasibi and Balog's ELR model \cite{hasibi2016exploiting}. It is defined as follows:

\begin{equation} \label{rsdm_der}
\resizebox{0.5 \columnwidth}{!} 
{
 $f^{ER}_S(D^{E_i},D^{R_{i-1,i}}) =  \left [   (1 - \alpha) f(D^{E_i},D^{R_{i-1,i}}) + \alpha \frac{ n(E_i)} { N^R}    \right ]$
}
\end{equation}

where the linear interpolation implements the Jelinek-Mercer smoothing method with $\alpha \in [0,1]$ and $f(D^{E_i},D^{R_{i-1,i}}) = \{0,1\}$ which measures if the entity $E_i$ represented in $D^{E_i}$ belongs to the relationship $R(E_{i-1},E_i)$ represented in $D^{R_{i-1,i}}$. The background model employs the notion of entity popularity within the collection of relationship documents. $n(D^{E_i})$ represents the number of relationship documents $D^R$ that contain the entity $E_i$ and $N^R$ represents the total number of relationship documents in the \textit{relationship} index.

For E-R queries with more than one relationship sub-query, we draw an edge between consecutive relationship documents within the ERDM graph. This edge creates a 3-clique containing two relationship documents and one entity document. The feature function $f^{RER}_S$ measures if a given entity $E_i$ is shared between consecutive relationship documents within the graph. We opted to define a simple binary function:

\begin{equation}
f^{ER}_S(D^{E_i},D^{R_{i-1,i}}, D^{R_{i,i+1}}) = 1 \ if \  E_i \in D^{E_i} \cap D^{R_{i-1,i}} \cap D^{R_{i,i+1}}\ , \ 0 \ \text{otherwise}
\end{equation}

In summary, we described the set of feature functions associated with each clique configuration within the ERDM graph. We leave for future work the possibility of exploring other type of features to describe textual similarity and compatibility between different nodes in the ERDM graph, such as neural language models.

\subsection{Ranking}

We have defined the set of clique configurations and the real valued feature functions that constitute the non-negative potential functions over the cliques in the graph of ERDM. We can now formulate the calculation of the posterior $P(D^E,D^R|Q$ using the probability mass function of the MRF, as follows:

\begin{align}
\begin{split}\label{eq:3}
   P_{\Lambda}(D^E,D^R|Q)  {}  \stackrel{\text{$rank$}}{=} &  \sum_{c \in C(G)} \lambda_c f(c)\\
     \stackrel{\text{$rank$}}{=} & \lambda^{E}_T \sum_E \sum_{Q^{E_i}} f^{E}_T(q^{E_i}_j,D^{E_i}) +\\
& \lambda^{E}_O \sum_E \sum_{Q^{E_i}} f^{E}_O(q^{E_i}_j,q^{E_i}_{j+1},D^{E_i}) +\\
& \lambda^{E}_U \sum_E \sum_{Q^{E_i}} f^{E}_U(q^{E_i}_j,q^{E_i}_{j+1},D^{E_i}) +\\
& \lambda^{R}_T \sum_R \sum_{Q^{R_{i,j}}}  f^{R}_T(q^{R_{i-1,i}}_j,D^{R_{i-1,i}}) + \\
& \lambda^{R}_O \sum_R \sum_{Q^{R_{i,j}}} f^{R}_O(q^{R_{i-1,i}}_j,q^{R_{i-1,i}}_{j+1} ,D^{R_{i-1,i}}) +\\
& \lambda^{R}_U \sum_R \sum_{Q^{R_{i,j}}} f^{R}_U(q^{R_{i-1,i}}_j,q^{R_{i-1,i}}_{j+1} ,D^{R_{i-1,i}}) +\\
& \lambda^{ER}_S \sum_R \sum_{E} f^{ER}_S(D^{E_i},D^{R_{i-1,i}})+  \\
& \lambda^{RER}_S \sum_R \sum_{E} f^{ER}_S(D^{E_i},D^{R_{i-1,i}},D^{R_{i,i+1}})  \\
\end{split}\\
\end{align}

In essence, E-R retrieval using the ERDM corresponds to ranking candidate entity tuples using a linear weighted sum of the feature functions over the cliques in the graph. Therefore, we can apply any linear learning to rank algorithm to optimize the ranking with respect to the vector of feature weights $\Lambda$. Given a training set $\mathcal{T}$ composed by relevance judgments, a ranking of entity tuples $\mathcal{R}_{\Lambda}$ and an evaluation function $\mathcal{E}(\mathcal{R}_{\Lambda};\mathcal{T})$ that produces a real valued output, our objective is to find the values of the vector $\Lambda$ that maximizes $\mathcal{E}$. As explained in \cite{metzler2007linear}, we require $\mathcal{E}$ to only consider the ranking produced and not individual scores. This is the standard characteristic among information retrieval evaluation metrics (e.g. MAP or NDCG).

\subsection{Discussion}
In this section we introduced the Entity-Relationship Dependence Model (ERDM), a novel supervised Early Fusion-based model for E-R retrieval. Inspired by recent work in entity retrieval we believe that modeling term dependencies between sub-queries and entity/relationship documents can increase search performance. 

ERDM can be seen as an extension of the SDM model \cite{metzler2005markov} for ad-hoc document retrieval in a way that besides modeling query term dependencies we create graph structures that depict dependencies between entity and relationship documents. Consequently, instead of computing a single posterior $P(D|Q)$ we propose to use the MRF for retrieval for computing a joint posterior of multiple entity and relationship documents given a E-R query, $P(D^E,D^R|Q)$. 

Moreover, since ERDM is a supervised model, we believe that tuning weights of feature functions, besides optimizing search performance, can also help to explain the inter-dependencies between sub-query terms and the respective documents, but also how entity documents and relationship documents contribute to the overall relevance of entity tuples given a E-R query.

\section{Experimental Setup}

In this section we detail how we conducted our experiments in E-R retrieval. Since we only have access to test collections comprising general purpose E-R queries we decided to use a Web corpus as dataset, more precisely ClueWeb-09-B\footnote{\url{https://lemurproject.org/clueweb09/}}.The ClueWeb09 dataset was created to support research on information retrieval and related human language technologies and contains 1 billion web pages. The part B is a subset of the most popular 50 million English web pages, including the Wikipedia. Part B was created as a resource for research groups without processing power for processing the all ClueWeb09 collection. We used the ClueWeb-09-B Web collection with FACC1 text span annotations linked to Wikipedia entities to show how RELink can be used for E-R retrieval over Web content. We developed our prototype using Apache Lucene for indexing and search. We used a specific Python library (PyLucene) that allowed our customized implementation tailored for E-R retrieval.

\subsection{Data and Indexing}

\begin{figure}[h] 
\centering
\includegraphics[width=0.7\textwidth]{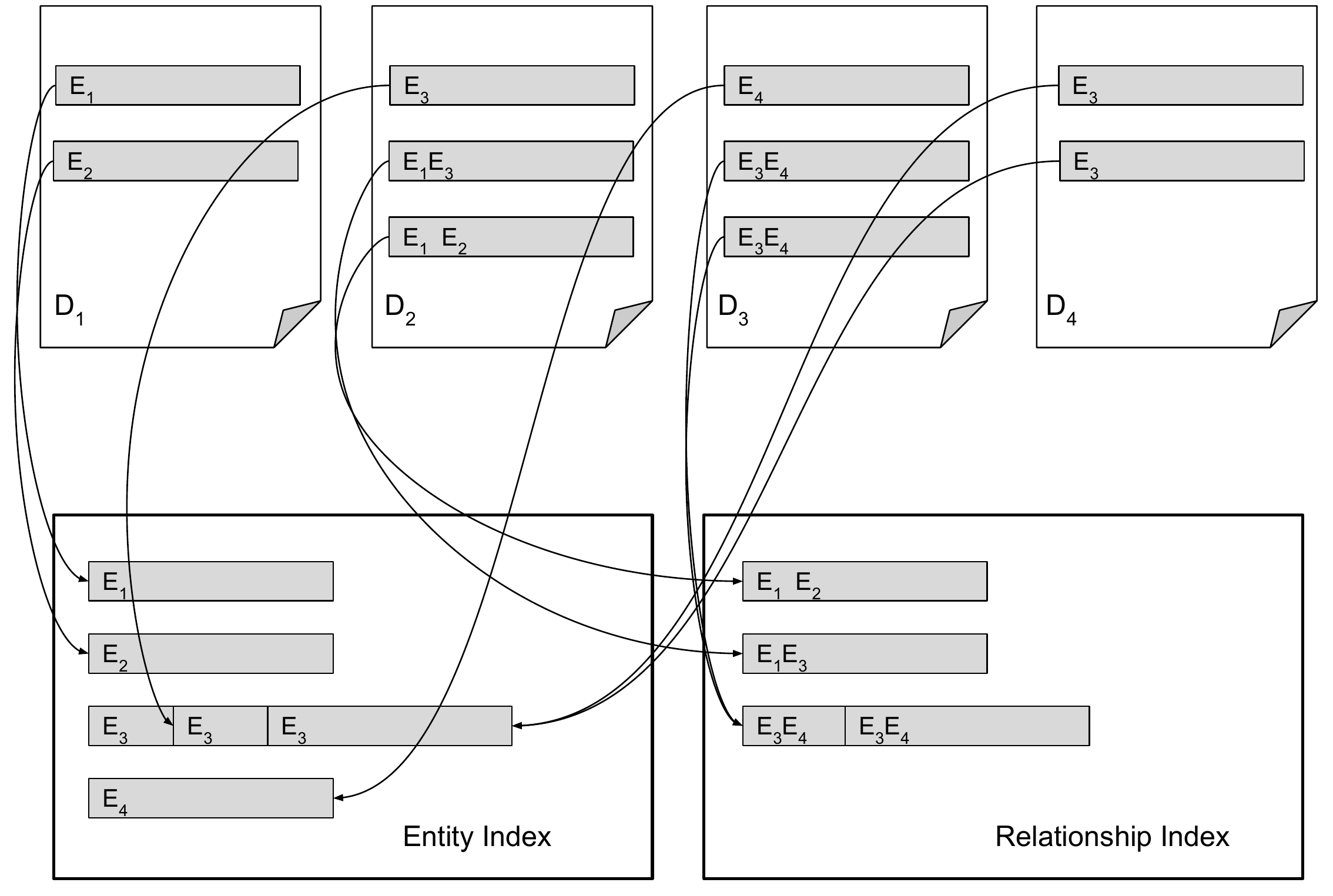}
\caption{Illustration of E-R indexing from a web corpus.}\label{ind}
\end{figure}

As a text corpus, we use ClueWeb-09-B combined with FACC1 text span annotations with links to Wikipedia entities (via Freebase). The entity linking precision and recall in FACC1 is estimated to be 80-85\% and 70-85\%, respectively \cite{gabrilovich2013facc1}. For our experiments we created two main indexes: one for entity extractions and one for entity pairs (relationships) extractions. We extract entity and pairs occurrences using an Open Information Extraction method like OLLIE \cite{schmitz2012open} over the annotated ClueWeb-09-B corpus as follows. For each entity annotation, we extract the sentence where it occurred as an entity context. For pairs of entities, we look for co-occurring entities in the same sentence and we extract the separating string, i.e., the context of the relationship connecting them. Figure \ref{ind} illustrates the indexing process adopted in this work. 

We obtained 476 million entity extractions and 418 million entity pairs extractions, as described in Table \ref{extractions}. In order to compute $|D^{E_i}|$ and $|D^{R_{i-1,i}}|$ we incrementally updated two auxiliary indices, containing the number of terms per entity and per entity pair, respectively. We ran our experiments using Apache Lucene and made use of GroupingSearch for grouping extractions by entity and entity pair at query time. To get the statistics for ordered and unordered bigrams we made use of SpanNearQuery.

\begin{table}[h]
\caption{ClueWeb09-B extractions statistics.}
\centering
\setlength\extrarowheight{1pt}
\begin{tabular}{l|l|l|l|}
\cline{2-4}
              & Total  & Unique     & Avg. doc. len.   \\ \cline{1-4}  
\multicolumn{1}{|l|}{Entities}     & 476,985,936 & 1,712,010  & 9977 \\     \cline{1-4}
\multicolumn{1}{|l|}{Entity pairs} & 418,079,378 & 71,660,094 & 138   \\ \cline{1-4}
\end{tabular}
\label{extractions}
\end{table}

\subsection{Retrieval Method and Parameter Tuning}

For experiments using ERDM we adopted a three stage retrieval method. First, queries $Q^{E_{i-1}}$,$Q^{E_i}$ are submitted against the entity index and $Q^{R_{i-1,i}}$ is submitted against the entity-pair index. Initial sets of top 20000 results grouped by entity or entity-pairs, respectively, are retrieved using Lucene's default search settings. Second, the feature functions of the specific retrieval model are calculated for each set, using an in-house implementation. This process is easily parallelized. The final ranking score for each entity-pair is then computed using the learned $\lambda$ weights. Evaluation scores are reported on the top 100 entity-pair results.

Parameter tuning for ERDM and baselines was directly optimized with respect to the Mean Average Precision (MAP). We make use of the RankLib's implementation of the coordinate ascent algorithm under the sum normalization and non-negativity constraints with 3 random restarts. Coordinate ascent is a commonly used optimization technique \cite{metzler2007linear} that iteratively optimizes a single parameter while holding all other parameters fixed.

Parameters are estimated using 5-fold cross validation for each of the 4 query sets separately. To be able to use the same train and test folds throughout all experiments, we first randomly create fixed train and test folds from the initial result set, for each query set. All reported evaluation metrics were macro-averaged over 5 folds.

We do not optimize the Dirichlet priors $\mu^E$ and $\mu^R$ in language models and set them equal to the traditional average document length, i.e., the average entity and entity pairs extractions length, respectively. The unordered window size $N$ for $f^{E}_U$ and $f^{R}_U$ is set to be 8, as suggested in \cite{metzler2005markov}. 

\subsection{Test Collections}
We ran experiments with a total of 548 E-R queries. We decided to just perform experiments using queries aiming 2-tuples of entities. We leave for future work the evaluation of queries aiming at triples. Besides RELink QC we used other 3 relationship-centric query sets, with pairs of Wikipedia entities as answers, i.e., relevance judgments. The query sets cover a wide range of domains as described in Table \ref{query_table}.  Query sets for entity-relationship retrieval are scarce. Generally entity retrieval query sets are not relationship-centric \cite{yahya2016relationship}.

\begin{table}[h]
\caption{Description of query sets used for evaluation.}
\centering
\setlength\extrarowheight{1pt}
\begin{tabular}{|l|l|p{7cm}|}
\cline{1-3}
\small Query Set  & Count & Domains   \\ \cline{1-3}
\small QALD-2     & 79     &   Geography and places, Politics and society, Culture and the Arts, Technology and science           \\ \cline{1-3}
 \small ERQ        & 28     &  \small Award, City, Club, Company, Film, Novel, Person, Player, Song, University                \\ \cline{1-3}
\small COMPLEX    & 60     & \small Cinema, Music, Books, Sports, Computing, Military conflicts                     \\ \cline{1-3}
\small RELink & 381    &  \small General Reference, Culture and the Arts, Geography and places, Mathematics and logic, Natural and physical Sciences, People, Religion and belief systems, Society and social sciences, Technology and applied science                    \\ \cline{1-3}
Total      & 548    &             \\ \cline{1-3}
\end{tabular}
\label{query_table}
\end{table}

One exception is the QALD-2 query set used in the DBpedia-entity collection \cite{balog2013test}. It contains a subset of relational queries, e.g.``\textit{Who designed the Brooklyn Bridge?}''. Most of relational queries in QALD-2 have a fixed relevant entity, e.g., ``\textit{Brooklyn Bridge}'' and can be easily transformed from single entity relevance judgments into pairs. From the 79 relational queries in QALD-2, we identified 6 with no fixed relevant entity in the query (e.g. ``\textit{Give me the capitals of all countries in Africa.}''). In these cases, for provided single entity relevance judgment we needed to annotate the missing entity manually to create a pair. For instance, given a capital city in Africa we identified the corresponding African country.

In addition, we used two benchmarks created in previous work using Semantic-Web-based approaches: ERQ \cite{li2012entity} and COMPLEX \cite{yahya2016relationship}. Neither ERQ nor COMPLEX provide complete relevance judgments and consequently, we manually evaluated each answer in our experiments. 
ERQ consists of 28 queries that were adapted from INEX17 and OWN28 \cite{li2012entity}. However, 22 of the queries have a given fixed entity in the query (e.g. \textit{``Find Eagles songs''}). Only 6 queries are asking for pairs of unknown entities, such as ``\textit{Find films starring Robert De Niro and please tell directors of these films.}''. 

COMPLEX queries were created with a semi-automatic approach \cite{yahya2016relationship}. It contains 70 queries from which we removed 10 that expect 3-tuples of entities. This query set consists of pure relationship-centric queries for unknown pairs of entities, such as ``\textit{Currency of the country whose president is James Mancham} ``\textit{Kings of the city which led the Peloponnesian League.}''  and ``\textit{Who starred in a movie directed by Hal Ashby?}''.

We used four different retrieval metrics, Mean Average Precision at 100 results (MAP), precision at 10 (P@10), mean reciprocal rank (MRR) and normalized discounted cumulative gain at 20 (NDCG@20). 


\section{Results and Analysis}

We start by performing a simple experiment for comparing Early Fusion and ERDM using both Language Models (LM) and BM25 as retrieval functions. Since we are only interested in comparing relative performance we opted to scale down our experimental setup. Instead of computing the term frequency for every extraction for a given entity or relationship we cap to 200 the number for each group of documents retrieved in the first passage. We tried several different values and for values below 200 extraction the performance reduced significantly. For 200, while the performance reduces it is not dramatic. This setup reduces the experimental runtime and since we had limited resources this proved to be useful. 

Table \ref{eferdm} depicts the results for this comparative evaluation. We decided to only use the three test collections specifically tailored for relationship retrieval. As we can see the results are very similar between EF and ERDM for both LM and BM25 variants. In the three test collections ERDM presents slightly better performance than the corresponding EF variant (e.g. BM25). However when performing statistical significance tests we obtained p-values above 0.05 when comparing EF and ERDM. This is very interesting as it shows that for general purpose E-R evaluation the overhead of computing sequential dependencies does not carry significant improvements.

\begin{table}[h]
\centering
\caption{Early Fusion and ERDM comparison using LM and BM25.}
\label{eferdm}
\begin{tabular}{|l|l|l|l|l|}
\hline
          & \multicolumn{4}{c|}{\textbf{ERQ}}     \\ \hline
          & MAP  & P@10    & MRR    & NDCG@20 \\ \hline
EF-LM     & \textbf{0.251}    & 0.15    & 0.3408 & 0.3508  \\ \hline
EF-BM25   & 0.1939   & 0.1423  & 0.1783 & 0.2861  \\ \hline
ERDM-LM   & \textbf{0.2611}   & 0.1615  & 0.3151 & 0.3589  \\ \hline
ERDM-BM25 & 0.2106   & 0.1462  & 0.2839 & 0.3257  \\ \hline
          & \multicolumn{4}{c|}{\textbf{COMPLEX}} \\ \hline
          & MAP  & P@10    & MRR    & NDCG@20 \\ \hline
EF-LM     & 0.1703   & 0.0596  & 0.1839 & 0.2141  \\ \hline
EF-BM25   & \textbf{0.1855}   & 0.0719  & 0.1907 & 0.2454  \\ \hline
ERDM-LM   & 0.1719   & 0.0789  & 0.2466 & 0.2492  \\ \hline
ERDM-BM25 & \textbf{0.1955}   & 0.0772  & 0.2257 & 0.248   \\ \hline
          & \multicolumn{4}{c|}{\textbf{RELink(381 queries)}}  \\ \hline
          & MAP  & P@10    & MRR    & NDCG@20 \\ \hline
EF-LM     & 0.0186   & 0.0063  & 0.0192 & 0.0249  \\ \hline
EF-BM25   & \textbf{0.0203}   & 0.0071  & 0.0227 & 0.0259  \\ \hline
ERDM-LM   & 0.0213   & 0.0058  & 0.0273 & 0.0255  \\ \hline
ERDM-BM25 & \textbf{0.0213}   & 0.0061  & 0.0265 & 0.0275  \\ \hline
\end{tabular}
\end{table}

On the other hand, we detect sensitivity to the retrieval function used. In ERQ, both ERDM-LM and EF-LM outperform BM25 but the opposite happens for COMPLEX and RELink. This sensitivity means that we cannot generalize the assumption that one of the retrieval functions is more adequate for E-R retrieval. 

Another important observation has to do with the overall lower results on the RELink test collection in comparison with ERQ and COMPLEX. Contrary to our expectations ClueWeb-09B has very low coverage of entity tuples relevant to the RELink test collection. 

We now present the results of comparing ERDM with three baselines using sequential dependence to evaluate the impact of modeling dependencies between query terms. The first baseline method, BaseEE, consists in submitting two queries against the entity index: $Q^{E_{i-1}} + Q^{R_{i-1,i}}$ and $Q^{R_{i-1,i}} + Q^{E_i}$. Entity-pairs are created by cross product of the two entity results set retrieved by each query. For each method we compute the Sequential Dependence Model(SDM) \cite{metzler2005markov} scores.

The second baseline method, BaseE, consists in submitting again a single query $Q$ towards the entity index used in ERDM. Entity-pairs are created by cross product of the entity results set with itself. The third baseline method, BaseR, consists in submitting a single query $Q$ towards an entity-pair index. This index is created using the full sentence for each entity-pair co-occurrence in ClueWeb-09-B, instead of just the separating string as in ERDM. This approach aims to capture any entity context that might be present in a sentence. ERDM relies on the entity index for that purpose.

In this evaluation we decided to not cap the number of extractions to compute term frequencies inside each group of results returned from the first passage with Lucene GroupingSearch. Due to the low coverage of ClueWeb for the entire RELink collection, we decided to just perform the evaluation using the top 100 queries with highest number of relevance judgments in our indexes. We also include results for the adapted QALD-2 test collection.

\begin{table}[h]
\centering
\caption{Results of ERDM compared with three baselines.}
\label{res}
\begin{tabular}{|l|l|l|l|l|}
\hline
          & \multicolumn{4}{c|}{\textbf{QALD-2}}           \\ \hline
          & MAP    & P@10     & MRR     & NDCG@20 \\ \hline
BaseEE & 0.0087 & 0.0027   & 0.0093  & 0.0055  \\ \hline
BaseE  & 0.0306 & 0.004684 & 0.0324  & 0.0363  \\ \hline
BaseR  & 0.0872 & 0.01678  & 0.0922  & 0.0904  \\ \hline
ERDM      & \textbf{0.1520} & 0.0405   & 0.1780  & 0.1661  \\ \hline
          & \multicolumn{4}{c|}{\textbf{ERQ}}              \\ \hline
          & MAP    & P@10     & MRR     & NDCG@20 \\ \hline
BaseEE & 0.0085 & 0.004    & 0.00730 & 0.0030  \\ \hline
BaseE  & 0.0469 & 0.01086  & 0.0489  & 0.038   \\ \hline
BaseR  & 0.1041 & 0.05086  & 0.1089  & 0.1104  \\ \hline
ERDM      & \textbf{0.3107} & 0.1903   & 0.37613 & 0.3175  \\ \hline
          & \multicolumn{4}{c|}{\textbf{COMPLEX}}          \\ \hline
          & MAP    & P@10     & MRR     & NDCG@20 \\ \hline
BaseEE & 0.0035 & 0        & 0.00430 & 0       \\ \hline
BaseE  & 0.0264 & 0.005    & 0.03182 & 0.1223  \\ \hline
BaseR  & 0.0585 & 0.01836  & 0.0748  & 0.0778  \\ \hline
ERDM      & \textbf{0.2879} & 0.1417   & 0.32959 & 0.3323  \\ \hline
          & \multicolumn{4}{c|}{\textbf{RELink(100 queries)}}           \\ \hline
          & MAP    & P@10     & MRR     & NDCG@20 \\ \hline
BaseEE & 0.03   & 0.01     & 0.0407  & 0.02946 \\ \hline
BaseE  & 0.0395 & 0.019    & 0.0679  & 0.03948 \\ \hline
BaseR  & 0.0451 & 0.021    & 0.0663  & 0.07258 \\ \hline
ERDM      & \textbf{0.1249} & 0.048    & 0.1726  & 0.1426  \\ \hline
\end{tabular}
\end{table}

Table \ref{res} presents the results of our experiments on each query set. We start by comparing the three baselines among each other. As follows from Table \ref{res}, BaseR baseline outperforms BaseEE and BaseE on all query sets, while BaseEE is the worst performing baseline. The BaseR retrieval is the only relationship-centric approach from the three baselines, as its document collection comprises entity-pairs that co-occurred in ClueWeb-09-B corpus. BaseEE and BaseE retrieve entity pairs that are created in a post-processing step which reduces the probability of retrieving relevant results. This results shows the need for a relationship-centric document collection when aiming to answer entity-relationship queries.

ERDM significantly outperform all baselines on all query sets. We performed statistical significance testing of MAP using ERDM against each baseline obtaining p-values below 0.05 on all the query sets. This results show that our Early Fusion approach using two indexes (one for entities and other for relationships) is adequate and promising. We believe this approach can become a reference for future research in E-R retrieval from an IR-centric perspective.

Nevertheless, based on the absolute results obtained on each evaluation metric and for each query set we can conclude that E-R retrieval is still very far from being a solved problem. There is room to explore new feature functions and retrieval approaches. This is a very difficult problem and the methods we proposed are still far from optimal performance. Queries such as ``\textit{Find world war II flying aces and their services}'' or ``Which mountain is the highest after Annnapurna?'' are examples of queries with zero relevant judgments returned. 

On the other hand, ERDM exhibits interesting performance in some queries with high complexity, such as ``Computer scientists who are professors at the university where Frederick Terman was a professor.'' We speculate about some aspects that might influence performance. 

One aspect has to do with the lack of query relaxation in our experimental setup. The relevant entity tuples might be in our indexes but if the query terms used to search for entity tuples do not match the query terms harvested from ClueWeb-09B it is not possible to retrieve those relevant judgments. Query relaxation approaches should be tried in future work. More specifically, with the recent advances in word embeddings it is possible to expand queries with alternative query terms that are in the indexes.

On the other hand, we adopted a very simple approach for extracting entities and relationships. The use of dependency parsing and more complex methods of relation extraction would allow to filter out noisy terms. We also leave this for future work. Moreover, to further assess the influence of the extraction method we propose to use selective text passages containing the target entity pairs and the query terms associated as well. Then different extraction methods could be tried and straightforward evaluation of their impact.

\begin{figure*}[h]
    \centering
    \begin{subfigure}[t]{0.5\textwidth}
        \centering
        \includegraphics[width= 1.0\linewidth]{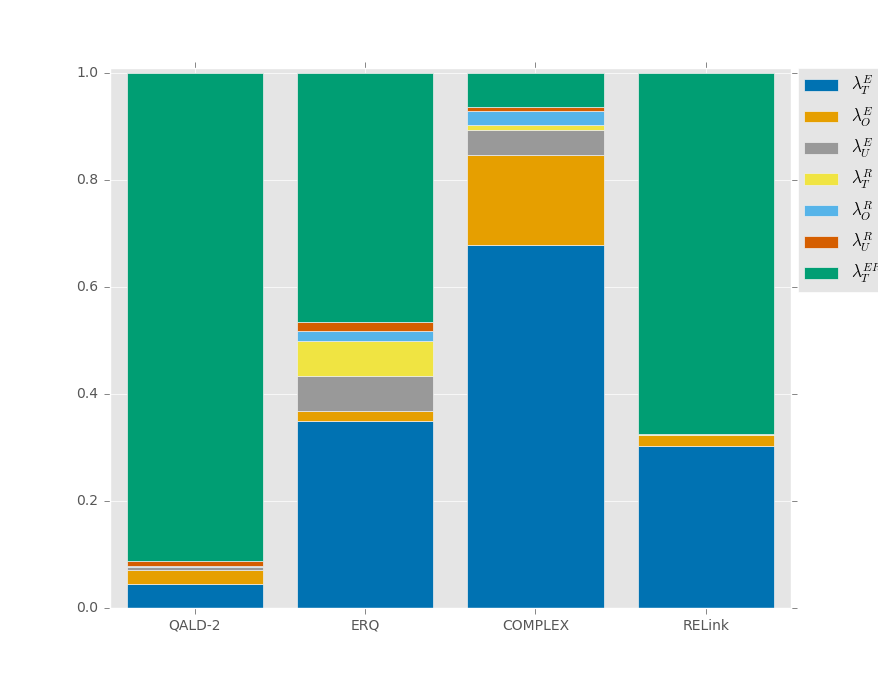}
        \caption{}
    \end{subfigure}%
    ~ 
    \begin{subfigure}[t]{0.5\textwidth}
        \centering
        \includegraphics[width= 1.0 \linewidth]{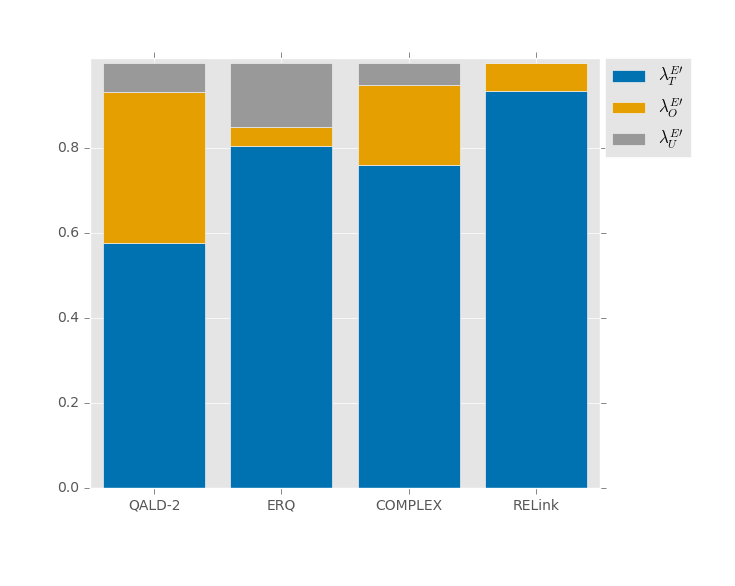}
        \caption{}
    \end{subfigure}

    \begin{subfigure}[t]{0.5\textwidth}
    \centering
    \includegraphics[width = 1.0\linewidth]{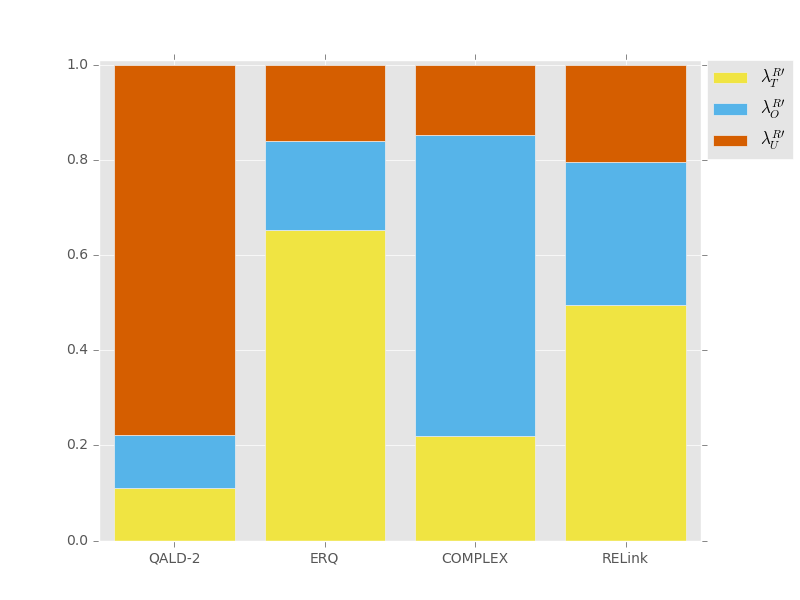}
    
          \caption{}
    \end{subfigure}
    \caption{Values of $\lambda$ for ERDM: (a) all $\lambda$, (b) $\lambda^{E'}$, (c) $\lambda^{R'}$. (b) and (c) were obtained using sum normalization. }
    \label{lambdas}
\end{figure*}

To understand how much importance is attributed to the different types of clique sets, we plot the values of the lambda parameters:  $\lambda^E$ parameters represent the feature importance of the set of functions targeting the dependence between entity query terms and the entity documents in overall ranking score for entity-pairs; $\lambda^R$ represent the importance of the feature functions of the relationship type queries and finally, the value for $\lambda^{ER}$ which is assigned to the feature function that evaluates if each entity retrieved from both entity type queries belongs to the entity-pair retrieved from the relationship type query. 

We plot the feature weights learned on each query set, as depicted in Figure \ref{lambdas}. We see that $\lambda^{ER}$  and $\lambda^{E}_{T}$ (weight for the unigram language model in the entity type queries) dominate the ranking function. We further evaluated the relative weights for each one of the three SDM-like functions using a sum normalization of the three weights for both entity documents and entity-pair documents. We observe that $\lambda^{E}_{T}$ dominates on every query set, however the same does not happen with $\lambda^{R}_{T}$. For relationship type queries the bigram features have higher values for COMPLEX and RELink.

\section{Conclusions}
Entity-Relationship (E-R) Retrieval is a complex case of Entity Retrieval where the goal is to search for multiple unknown entities and relationships connecting them. Contrary to entity retrieval from structured knowledge graphs, IR-centric approaches to E-R retrieval are more adequate in the context of ORM. This happens due to the dynamic nature of the data sources which are much more transient than other more stable sources of information (e.g Wikipedia) used in general Entity Retrieval. Consequently, we developed E-R retrieval methods that do not rely on fixed and predefined entity types and relationships, enabling a wider range of queries compared to Semantic Web-based approaches.

We started by presenting a formal definition of E-R queries where we assume that a E-R query can be decomposed as a sequence of sub-queries each containing keywords related to a specific entity or relationship. Then we adopted a probabilistic formulation of the E-R retrieval problem. When creating specific representations for entities (e.g. context terms) and for pairs of entities (i.e. relationships) it is possible to create a graph of probabilistic dependencies between sub-queries and entity plus relationship representations. We use a Bayesian network to depict these dependencies in a probabilistic graphical model. To the best of our knowledge this represents the first probabilistic model of E-R retrieval.  

However, these conditional probabilities cannot be computed directly from raw documents in a collection. In fact, this is a condition inherent to the problem of Entity Retrieval. Documents serve as proxies to entities and relationship representations and consequently, we need to fuse information spread across multiple documents to be able to create those representations. We proposed to tackle E-R Retrieval using Early Fusion, inspired from Balog\'s Model 1 \cite{balog2006formal}. Early Fusion for E-R Retrieval aggregates context terms of entity and relationship occurrences to create two dedicated indexes, the \textit{entity} index and the \textit{relationship} index. Once we have the two indexes it is possible to apply any retrieval method to compute the relevance scores of entity and relationship documents (i.e. representations) given the E-R sub-queries. The joint probability to retrieve the final entity tuples is computed using a factorization of the conditional probabilities, i.e., the individual relevance scores.

Furthermore, we developed a novel supervised Early Fusion-based model for E-R retrieval, the Entity-Relationship Dependence Model (ERDM). It uses Markov Random Field to model term dependencies of E-R sub-queries and entity/relationship documents. ERDM can be seen as an extension of the Sequential Dependence Model (SDM) \cite{metzler2005markov} for ad-hoc document retrieval in a way that it relies on query term dependencies but creates a more complex graph structure that connects terms of multiple (sub-)queries and multiple documents to compute the probability mass function under the MRF.

We performed experiments at scale using the ClueWeb-09B Web corpus from which we extracted and indexed more than 850 million entity and relationship occurrences. We evaluated our methods using four different query sets comprising a total of 548 E-R queries. As far as we know, this is the largest experiment in E-R retrieval, considering the size of the query set and the data collection. Results show consistently better performance of the ERDM model over three proposed baselines. When comparing Language Models and BM25 as feature functions we observed variance on the performance depending on the query set. Furthermore, using unsupervised Early Fusion proved to be very competitive when compared to ERDM, suggesting that it can be used in some application scenarios where the overhead of computing sequential dependencies might be unfeasible.

One of the most interesting avenues we would like to explore would be the use of neural networks as feature functions of the ERDM model. Since we have a dataset of more than 850 million entity and relationship extractions this represents an ideal scenario for Deep Learning. We propose to use a window based prediction task similar to the CBOW model for training word embeddings. Given a fixed window size, one would learn a neural network that would provide a ranked list of entities/relationships given an input query. We believe this approach would reduce the computational costs of the current ERDM feature functions since we would not need to keep two huge indexes at query time.

We would like also to explore different priors in entity and relationship documents within ERDM. For instance, creating source and time sensitive rankings would be useful when using transient information sources.

\bibliographystyle{ACM-Reference-Format}
\bibliography{references}   

\end{document}